\documentclass[12pt]{article}
\usepackage{jheppub}
\usepackage{graphics, color,soul}
\usepackage{graphicx} 
\usepackage{subcaption}
\usepackage{latexsym}
\usepackage{epsfig,amssymb,euscript,slashed}
\usepackage{amsmath}
\usepackage{xcolor}
\usepackage{tikz-cd} 
\usepgfmodule{shapes}
\usetikzlibrary{backgrounds, arrows,calc,shapes,decorations.pathreplacing, decorations.markings, decorations.pathmorphing, automata,positioning,calligraphy}

\tikzset{snake it/.style={decorate, decoration=snake}}

\tikzset{
  arrow/.pic={\path[tips,every arrow/.try,->,>=#1] (0,0) -- +(.1pt,0);},
  pics/arrow/.default={latex,very thick}
}

\begin{document}
\title{ 
The black hole behind the cut
}

\author{Stefano Giusto$^{a, b}$,}
\author{Cristoforo Iossa$^{c,d}$, and}
\author{Rodolfo Russo$^{e}$}

\affiliation{$^a$Dipartimento di Fisica, Università di Genova, Via Dodecaneso 33, 16146, Genoa, Italy.}
\affiliation{$^b$ I.N.F.N. Sezione di Genova, Via Dodecaneso 33, 16146, Genoa, Italy.}
\affiliation{$^c$SISSA, via Bonomea 265, 34136 Trieste, Italy}
\affiliation{$^d$INFN, sezione di Trieste, via Valerio 2, 34127 Trieste, Italy}
\affiliation{$^e$Centre for Theoretical Physics, Department of Physics and Astronomy,
Queen Mary University of London, Mile End Road, London, E1 4NS, United Kingdom.}

\emailAdd{stefano.giusto@unige.it}
\emailAdd{ciossa@sissa.it}
\emailAdd{r.russo@qmul.ac.uk}

\abstract{
  We study the analytic structure of the heavy-heavy-light-light holographic correlators in the supergravity approximation of the AdS$_3 \times S^3$/CFT$_2$ duality. As an explicit example, we derive the correlator where the heavy operator is a classical microstate of the 5D supersymmetric black hole and its dual geometry interpolates as a function of a continuous parameter between global AdS$_3$ and the extremal BTZ black hole. The simplest perturbation of this interpolating geometry by a light field is described by the Heun equation and we exploit the relation of its connection coefficients to the Liouville CFT to analytically compute the correlator in the two limits, focusing in particular on the black hole regime. In this limit we find that the real poles of the correlator become dense and can be approximated by a cut. We show that, when the charges of the heavy state are in the black hole regime, the discontinuity across the cut has complex poles corresponding to the quasi-normal modes of BTZ. This behaviour is qualitatively similar to what is expected for the large central charge limit of a typical black hole microstate.}
    
\maketitle

\section{Introduction}

The AdS/CFT duality~\cite{Maldacena:1997re} has provided a concrete way to count the microstates of black holes in terms of ``heavy'' operators in the dual Conformal Field Theory (CFT), {\rm i.e.} operators whose conformal dimension is of the order of the central charge $c$. This approach has been particularly successful for configurations that preserve some supersymmetries, see for instance the microscopic derivation by Strominger and Vafa~\cite{Strominger:1996sh} of the Bekenstein-Hawking entropy of certain extremal black holes. Building on this foundation, it is interesting to go beyond the counting problem and use the tools of the AdS/CFT duality to investigate the nature of individual microstates in the regime where semiclassical gravity is thought to be a good approximation to the full theory. In general this is a challenging problem since it requires to analyse a CFT at strong coupling and to deal with complicated operators. A natural question is how the general thermodynamic properties of the usual black hole solutions emerge from unitary CFT correlators involving pure states.

In this paper we focus on the issue above working in the framework of the AdS$_3$/CFT$_2$ duality that emerges in the decoupling limit of a stack of D1 and D5-branes in type IIB string theory compactified on $K3$ or $T^4$~\cite{Maldacena:1997re}. The observables of interest are the holographic correlators involving two heavy and two light states which are dubbed HHLL correlators: from the bulk point of view, the light operators correspond to elementary excitations of the supergravity fields that probe a non-trivial geometry representing the gravitational backreaction of the heavy operators. Ideally we would like to study such correlators involving typical states in the ensemble describing the Strominger-Vafa~\cite{Strominger:1996sh} or the BMPV black holes~\cite{Breckenridge:1996is}. However, since we do not know how to treat this problem analytically, we consider a simpler setup where the heavy states are atypical and correspond to coherent superposition of a large number ${\cal O}(c)$ of mutually BPS supergravity excitations (so they are often called ``graviton gas" states). On the bulk side, these states are described in terms of regular supergravity solutions that, in the AdS$_3$/CFT$_2$ context, are known as ``superstrata''~\cite{Bena:2015bea,Bena:2017xbt} (see~\cite{Shigemori:2020yuo} for a review). It is well known that the graviton gas family is not large enough to account for the black hole entropy~\cite{deBoer:1998us}, see~\cite{Shigemori:2019orj,Mayerson:2020acj} for a recent detailed discussion of this point. Because of this, one may discard the graviton gas and identify the black hole microstates with the remaining part of the heavy spectrum. However there are families of graviton gas microstates that in certain limit approach arbitrarily well the black hole regime in the following sense: in the gravitational picture they are described by geometries with a long throat which are well approximated by the standard black hole solution. We will focus exactly on such class of heavy states and in particular on the superstrata derived in~\cite{Bena:2016ypk}.

Thanks to AdS/CFT duality it is possible to calculate the HHLL correlators at strong coupling by studying the quadratic perturbations around the solutions dual to the heavy operators~\cite{Galliani:2017jlg,Bombini:2017sge,Bena:2019azk}. This quantity contains a large amount of dynamical information including the energies $\omega_n$ of the non-BPS excitations on the top of the supersymmetric heavy state. We will reassess the HHLL correlators studied in~\cite{Bena:2019azk} where the heavy states are the $(1,0,\tilde{n})$ superstrata~\cite{Bena:2016ypk} focusing on the analytic structure of such correlators. For $\tilde{n}>0$ these superstrata preserve the same supersymmetries of the three-charge black hole and are labelled by a continuous parameter, $\eta$, which sets the length of an extremal throat. When $\eta\ll 1$ the regular geometry dual to the heavy state are well approximated by the standard black hole solution and so, by studying this example, we expect to learn some lessons that are relevant for general black hole microstates. On the technical side, when $\tilde{n}=1,2$, the simplest perturbations around the supersymmetric solution are described by the Heun equation. 

Starting with~\cite{Nekrasov:2009rc}, several problems related to Heun equations have been solved exploiting a connection to Seiberg-Witten theory and its AGT dual: Liouville CFT \cite{Alday:2009aq}. These ideas have been recently applied in the context of black hole perturbations~\cite{Aminov:2020yma,Bonelli:2021uvf,Bonelli:2022ten,Bianchi:2021xpr,Bianchi:2021mft,Bianchi:2022qph,Bianchi:2023sfs,Consoli:2022eey}\footnote{See \cite{Fioravanti:2021dce,Fioravanti:2023zgi} for a similar approach based on integrability.} and holography \cite{Dodelson:2022yvn}, where quantities of interests are computed in terms of Nekrasov-Shatasvili (NS) partition functions. In particular in \cite{Bonelli:2022ten} the explicit connection coefficients of the Heun equations have been derived. Such connection coefficients are essential in our analysis, as the HHLL holographic correlators require to impose a regularity boundary condition in the interior of the geometry and another boundary condition (encoding the positions of the light operators) at the AdS boundary. A novelty of our analysis is that when $\eta\ll1$ the relevant Seiberg-Witten theory is in a strongly coupled phase. As a result, the connection coefficients in this regime are not computed by NS functions.

For any non-zero value of $\eta$ we find that the momentum-space HHLL correlators display a discrete set of real poles in the energy (or equivalently in the lightcone momenta~\eqref{eq:lightconem}) which is the expected behaviour for a unitary correlator. Notice that this is true in the $c\to\infty$ limit (where our supergravity analysis is reliable) hinting that the graviton gas states have a simple classical limit. By following the analysis of~\cite{Leutheusser:2021qhd,Leutheusser:2021frk,Leutheusser:2022bgi,Witten:2021jzq,Witten:2021unn,Chandrasekaran:2022eqq} this means that, in presence of the heavy states under consideration, the light states form a type I algebra even in the classical limit. However, if we extrapolate our results to the small $\eta$ regime, we see that the distance between consecutive poles goes to zero and one can effectively approximate the poles with a branch cut\footnote{To be precise we cannot trust the supergravity approach in the strict $\eta\to 0$ limit. Even if one takes the string length to be small, the expectation from the JT gravity is that this approximately AdS$_2$ solutions contain at least a strongly coupled mode that will effectively cut the length of throat (which classically is proportional to $|\ln\eta|$), see for instance~\cite{Lin:2022rzw,Lin:2022zxd} and references therein. In this paper, we will not study this interesting regime, as for our purposes it is sufficient that the separation between consecutive poles (set by $\eta$) is small with respect to their position: in that case we can say that the correlator effectively develops the branch cut mentioned above.}. By doing this, we approximate the spectrum of excitations around the heavy state with a continuum and so move from a type I to a type III algebra. 

When $\tilde{n}>0$, there is a bound of order one on $\eta$ (see Eq.~\eqref{eq:bhregeta}) below which the geometry dual to the heavy state becomes similar to that of extremal BTZ. The discontinuity along the cut that develops when the poles merge is directly related to the Wightman correlator in presence of the heavy state. This allows us to show that, in the small $\eta$ limit, the Wightman correlator for our HHLL configuration develops poles with a non-zero imaginary part that were absent in the exact result. More precisely the Wightman function takes exactly the functional form of the BTZ result~\cite{Birmingham:2001pj,Son:2002sd} and so the imaginary poles correspond to the BTZ QNMs. While our analysis is in the gravity limit and $\eta$ is the parameter that measures the deviation from the deep black hole regime, the pattern described above should apply also to the large $c$ limit of typical states. At finite $c$, we expect to have discrete real poles (even if possibly with an erratic behaviour as $c$ changes~\cite{Schlenker:2022dyo}), but at large $c$ the distribution of the poles should become dense and one could recover the black hole results by looking at the discontinuity along the emerge branch cut. This is what happens in our case in the small $\eta$ regime where the pole separation is small in comparison to their position. Of course the analogy is not perfect: for instance we cannot resolve the separation of all energy levels which is exponentially suppressed in the typical sector~\cite{Raju:2018xue}. Moreover a peculiarity of our three dimensional setup is that BTZ does not admit classically stable orbits and this implies that the poles in the complex plane always have a large imaginary part. In higher dimensions, the discontinuity along the branch cut develops also poles that are exponentially close to the real axis in the large spin limit~\cite{Dodelson:2022eiz}, so the analytic structure is even richer\footnote{Thus these poles can mimic a discrete spectrum when applying the light-cone bootstrap approach to HHLL correlators involving black hole geometries~\cite{Kulaxizi:2018dxo,Karlsson:2019qfi,Fitzpatrick:2019zqz,Kulaxizi:2019tkd,Li:2019zba,Li:2020dqm,Parnachev:2020zbr,Karlsson:2021mgg,Karlsson:2022osn}.}. In contrast in AdS$_3$ there are stable orbits for conical defects~\cite{Fitzpatrick:2014vua}, {\rm i.e.} for states below the BTZ threshold. This case was studied in detail in~\cite{Kulaxizi:2018dxo,Karlsson:2019qfi,Fitzpatrick:2019zqz,Kulaxizi:2019tkd,Li:2019zba,Li:2020dqm,Giusto:2020mup} where the heavy-light spectrum obtained from a bootstrap approach was related to the phase shift of the probe in the dual geometry.

The structure of this paper is the following. In Section~\ref{sec:state-dualg} we introduce the general AdS$_3$/CFT$_2$ setup relevant for the rest of the paper. In particular we provide a brief description of the states we use, considering both the CFT and the bulk point of view. We then derive the equation encoding the simplest HHLL correlator for the $(1,0,\tilde{n})$ superstrata which will be the main character in our discussion. In Section~\ref{sec:strongcorr}, we study this equation with the boundary conditions relevant for the holographic correlators. We first recall the derivation of~\cite{Bombini:2017sge} for the case $\tilde{n}=0$ which reduces to the standard hypergeometric equation. Then we move on to the case $\tilde{n}=1$ where one obtains a reduced confluent Heun equation and pay particular attention to the limits $\eta\to 1$ (which is a small deformation of AdS studied with a perturbative approach in~\cite{Bombini:2019vnc}) and $\eta\to 0$ (where one recovers the extremal BTZ solution). In Section~\ref{sec:anal-str} we study the analytic structure of the momentum space correlator again starting from the $\tilde{n}=0$ case as a warming up exercise and then moving to $\tilde{n}=1$. In the latter case, we show how the BTZ result is reproduced when considering the discontinuity along the branch cut emerging at small $\eta$. In Section~\ref{sec:configuration} we discuss the configuration space formulation of the HHLL correlator focusing on the near lightcone limit. As a consistency check we show that it admits an sensible OPE expansion in terms of heavy-light Virasoro blocks~\cite{Fitzpatrick:2014vua,Fitzpatrick:2015zha}. We present our conclusion and outlook in Section~\ref{sec:concl} and collect some technical details in the appendices. Appendix~\ref{app:conn} provides some background material on how Liouville correlators are used to obtain information on the Heun connection formulae, in Appendix~\ref{app:BTZ} we give a short summary of the extremal BTZ correlator and in Appendix~\ref{sec:wkb} we perform a brief WKB analysis of the correlators in two special cases since this is useful to cross-check some details of the general results.

\section{CFT states and their dual geometries}\label{sec:state-dualg}
In this article we compute a holographic correlator in the tree-level approximation of supergravity, which corresponds to the large central charge and strong coupling limits of the dual CFT. Despite working in a classical effective field theory approximation, some crucial features of our holographic construction, like the intrinsic six-dimensional nature of the geometries dual to the heavy states, are determined by the fact that our effective theory derives from a UV complete theory of quantum gravity, namely type IIB superstring theory compactified on a 4-dimensional Calabi-Yau space $\mathcal{M}$, which could be either $T^4$ or $K3$, times $S^1$. We will thus start by reviewing both our gravity setting and the dual CFT, which is known as the D1-D5 CFT.

For the states we consider, the geometry of the compact 4D space $\mathcal{M}$ is trivial up to an overall volume factor and we can thus focus on the remaining six dimensions, where we have a $\mathcal{N}=(2,2)$ supergravity theory with $N_f=5$ tensor multiplets for $\mathcal{M}=T^4$, or a $\mathcal{N}=(2,0)$ theory with $N_f=21$ tensor multiplets for $\mathcal{M}=K3$. In this article we concentrate on the near-horizon limit where the 6D geometries have AdS$_3\times S^3$ asymptotics\footnote{The extension of the geometries to the $\mathbb{R}^{4,1}\times S^1$ flat asymptotic region is also known \cite{Bena:2017xbt}.} and they can be holographically mapped to ``heavy" states of the D1-D5 CFT whose conformal dimensions scale like the central charge $c$ for large $c$. We use ``light" states, with a conformal dimension of order 1 for $c\to\infty$, to probe the heavy states and we will, in particular, compute the HHLL correlator between two heavy and two light operators, by studying the linear fluctuations of the light fields in the heavy state background. 

In this section, we first recall the operator content of the D1-D5 CFT and introduce the heavy and light states relevant for our analysis, we summarise their dual gravity description and write down the wave equation that governs the linearised perturbation around the heavy state geometry.  

\subsection{The light states}\label{sec:lightstates}
The D1-D5 CFT has a chiral algebra generated by the supercharges, the SU(2) currents and the Virasoro generators\footnote{In $T^4$ case there are extra four holomorphic and four anti-holomorphic $U(1)$ currents.}. The only states that remain in the CFT spectrum in the supergravity limit are the chiral primary operators (CPO) and their chiral-algebra descendants. Each of the 6D tensor multiplets gives rise, upon reduction on the $S^3$, to a tower of CPO's, $s^{(f)}_k$, where $f=1,\ldots,N_f$ is the ``flavour" index and $k=1,\ldots,\infty$ parametrises the $S^3$ harmonics. An extra tower of CPOs, $\sigma_k$, comes from the gravity multiplet but now $k=2,\dots,\infty$. Both $s^{(f)}_k$ and $\sigma_k$ are spin 0 operators of conformal dimension $h=\bar{h}=\frac{m}{2}$. The gravity multiplet also generates spin 1 CPOs, $V^+_k$, with $k=0,\ldots,\infty$, of conformal dimensions $h=\bar{h}+1=\frac{k}{2}+1$, together with their anti-chiral counterparts, $V^-_k$; for $k=0$ these are just the SU(2) currents $J^\pm$. At the supergravity point the CFT has an $SO(N_f)$ symmetry acting on the flavour index $f$, which reflects on the structure of the correlators: for example a four-point correlator between the CPOs $s_k^{(f)}$ can be non-vanishing only if the flavour indices are pairwise equal.

On the gravity side these CPOs are dual to 6D fields that satisfy not-so-simple wave equations already in the AdS$_3\times S^3$ vacuum, and things become significantly more complicated in the non-trivial background of a heavy state. To probe heavy states it is computationally more convenient to consider the descendants obtained by acting on $s^{(f)}_k$ with one left-moving and one right-moving supercharge; schematically these operators are $G^- {\tilde G}^- s_k^{(f)}$. As these will represent our light operators in the remainder of this article, we will simply denote them as $O_L$, suppressing both the flavor index $f$ and the KK index $k$. Since a supercurrent $G^-$ ($\tilde G^-$) increases the left (right) conformal dimension by $\frac{1}{2}$ and decreases the left (right) $SU(2)$ charge by the same amount, $O_L$ has dimension $h=\bar h=\frac{k+1}{2}$ and charge $j=\bar j=\frac{k-1}{2}$. The lowest dimensional operators with $k=1$ are neutral scalars of dimension $(1,1)$, which are some of the moduli of the CFT; the operators with $k>1$ are higher KK modes of these scalars. In the AdS$_3\times S^3$ vacuum, $O_L$ is dual to a minimally coupled scalar in six dimensions. The linearised wave equation describing the perturbation dual to $O_L$ in the background of a non-trivial heavy state, is, in general, more complicated. We will see however that things simplify for particular heavy states.

\subsection{The 
heavy states and the geometries}
\label{sec:heavystate}

The class of heavy states that we use in our computation are multi-particle states made by many copies of the light states introduced above. We are interested in states that carry the same charges and left-moving angular momentum\footnote{Unlike the supersymmetric black holes, our states also carry some amount of right-moving angular momentum, that is controlled by a continuous parameter, $\eta$. In the $\eta\to 0$ limit, both the left and right-moving angular momenta vanish, and the solution reduces to the Strominger-Vafa black hole \cite{Strominger:1996sh}.} of a supersymmetric three-charge 5D black hole with a horizon of finite area (the BMPV black hole \cite{Breckenridge:1996is}). Having a third-charge requires that the constituent single-particle operators are descendants of CPO's obtained by acting with some left-moving chiral-algebra operator. Black holes with a finite-area horizon have a left-moving angular momentum that is not too big compared to the other charges: if we denote by $c=6 N$ the central charge, $n_P=h-\bar h$ the integer momentum charge and $j_R$ the left-moving angular momentum of the asymptotically flat solution (that is related to the $SU(2)$ charge of the CFT in the NS sector, $j$, by the spectral flow relation $j_R=j-\frac{N}{2}$), the condition for having a regular horizon is
\begin{equation}\label{blackholeregime}
N\,n_P-j_R^2>0\,.    
\end{equation}
To increase $n_P$ without increasing $j$ one can, for example, act on a CPO with the Virasoro generator\footnote{Acting with the generators $L_{-n}$ with $n>1$, that are outside the global $SL(2,\mathbb{R})$ sub-algebra, leads in general to stringy states.} $L_{-1}$. Moreover it turns out that choosing for the CPO the lowest KK mode of a tensor multiplet, $s_1^{(f)}$, leads to the simplest dual geometry and, in particular, to a separable wave equation \cite{Bena:2017upb}. We are thus lead to consider a state of the form $\left(L^{\tilde{n}}_{-1} s_1^{(f)}\right)^{N_1}$, for some non-zero integer $\tilde{n}$. When the number of single-particle constituents, $N_1$, is of order $c$, the former state is heavy and it should admit a dual gravitational description in classical supergravity. More precisely \cite{Skenderis:2006ah}, to have a classical geometric description one should consider a coherent-type superposition of multi-particle states\footnote{Note that, for the state that we consider in this article, the $p$-particle operator $\left(L^{\tilde{n}}_{-1}s_1^{(f)}\right)^p$ is not defined by taking the OPE limit of the single-particle operators $L^{\tilde{n}}_{-1}s_1^{(f)}$; this subtlety, which is explained for example in \cite{Ceplak:2021wzz}, will not be particularly relevant for our analysis, but one should take into account that with our definition the multi-particle heavy states are not orthogonal to the single-particle ones, and some extremal correlators do not vanish.}, parametrised by the continuous parameter $\eta$ (with $\eta\in[0,1]$):
\begin{equation}
\label{defOH}
O_H=\sum_{p=0}^N (1-\eta^2)^\frac{p}{2}\eta^{N-p}\,\left(L^{\tilde{n}}_{-1}s_1^{(f)}\right)^p\,.    
\end{equation}
For large $N$ the sum over $p$ in \eqref{defOH} is peaked over $\bar p=N(1-\eta^2)$, as one can understand by looking at the norms of the individual terms in the sum. Thus the average conformal dimensions, $h_H, \bar h_H$, and angular momenta, $j_H, \bar j_H$, of $O_H$ are
\begin{equation}
h_H=N\left(\tilde{n}+\frac{1}{2}\right)(1-\eta^2)\quad,\quad \bar h_H=\frac{N}{2}(1-\eta^2)\quad,\quad j_H=\bar j_H=\frac{N}{2}(1-\eta^2)\,.  
\end{equation}
Focusing only on $n_P=N\,\tilde{n}\,(1-\eta^2)$ and the left-moving angular momentum in the R-sector, $j_R=-\frac{N}{2}\eta^2$, we see that the constraint for having a black hole with the same charges \eqref{blackholeregime} is satisfied if
\begin{equation}\label{eq:bhregeta}
\eta^2< 2\sqrt{\tilde{n}}\,(\sqrt{\tilde{n}+1}-\sqrt{\tilde{n}})\,.  
\end{equation}
This shows that, as far as $\tilde n\not=0$, our family of states enters the black hole regime for small enough values\footnote{Even for $\tilde{n}=1$ this is not a stringent constraint as it is sufficient to have $\eta^2\lesssim 0.828$.} of $\eta$. For $\eta=0$ the state $O_H$ reaches a somewhat degenerate limit:
\begin{equation}
   O_H\stackrel{\eta\to 0}{\to} \left(L^{\tilde{n}}_{-1}s_1^{(f)}\right)^N\,.  
\end{equation}
While for generic values of $\eta$ ($0<\eta<1$) the state \eqref{defOH} contains two different types of elementary constituents ($p$ copies of the state $L^{\tilde{n}}_{-1}s_1^{(f)}$ and $N-p$ copies of the vacuum), for $\eta=0$ the state is made of $N$ copies of the same state and the sum over $p$ in \eqref{defOH} degenerates to a single summand. A similar phenomenon happens for $\eta=1$, when the state reduces to the NS vacuum. As a consequence of this degenerate limit, the expectation values of all the supergravity operators (beside the stress-tensor) computed in the state $O_H$ vanish at $\eta=0$. Consider for example the expectation value of the operator $s_1^{(f)}$ itself: when it acts on the state $O_H$ in \eqref{defOH}, it turns a vacuum copy into a non-trivial copy of type $L^{\tilde{n}}_{-1}s_1^{(f)}$ and thus, when both type of copies are present, its expectation value in the state $O_H$ does not vanish~\cite{Kanitscheider:2007wq,Giusto:2015dfa}; but this fails for $\eta=0$ (or $\eta=1$) and thus 
\begin{equation}
    \langle O_H|s_1^{(f)}|O_H\rangle\stackrel{\eta\to 0}{\to} 0\,.
\end{equation}
One can see that the same conclusion holds for any other CPO (or their descendants) and thus, from the point of view of supergravity, the state $O_H$ with $\eta=0$ is indistinguishable from the thermal ensemble, where all classical expectation values of non-trivial operators are also expected to vanish. For this reason, though the state $O_H$ is certainly a very atypical representative of the thermal ensemble characterised by the left-moving temperature of the supersymmetric three-charge black hole, its $\eta\to 0$ limit should share some properties of typical states. This fact is reflected in the dual geometry which, as $\eta$ goes to zero and for $\tilde n>0$, develops a long AdS$_2$ throat and reduces precisely to the Strominger-Vafa black hole for $\eta\to 0$. We will thus refer to the small $\eta$ limit as the black hole limit. In the opposite regime, $\eta\to 1$, the heavy state approaches the $SL(2,\mathbb{C})$ invariant vacuum and the geometry is a small perturbation of global AdS$_3\times S^3$, which is attained exactly for $\eta=1$. In this limit we expect the physics to be far from the black hole behaviour.

The supergravity solution dual to $O_H$ for generic values of $\eta$ was first found in \cite{Bena:2016ypk}. We concentrate here on the 6D Einstein metric, but the full solution contains also scalars and three-form fluxes. In the 6D geometry the $S^3$ directions, $\theta$, $\varphi_1$, $\varphi_2$, are non-trivially fibered over the AdS$_3$ ones, $\rho$, $\tau$, $\sigma$, and the Einstein metric can be written as\footnote{The coordinates and parameters used here are related to those of \cite{Bena:2016ypk} by
\begin{equation}
 \rho=\frac{r}{a}\\,\,,\,\,\tau=\frac{t}{R_y}\,\,,\,\,\sigma=\frac{y}{R_y}\,\,,\,\,\varphi_1+\tau=\phi\,\,,\,\,\varphi_2+\sigma=\psi\,\,,\,\,\eta=\frac{a}{\sqrt{a^2+\frac{b^2}{2}}}\,.    
\end{equation}
Note that in this article we mostly work in the NS sector, while the states in \cite{Bena:2016ypk} where expressed in the R sector; the shifts of the $S^3$ coordinates $\varphi_i$ realise the spectral flow transformation between the two sectors.}
\begin{equation}\label{6DEinstein}
\begin{aligned}
ds^2_6=\frac{\Lambda}{G}ds^2_3+\Lambda d\theta^2&+\frac{\sin^2\theta}{\Lambda}\left(d\varphi_1+(1-\eta^2)d\tau\right)^2+ \\ &+\frac{G\cos^2\theta}{\Lambda}\left(d\varphi_2+d\sigma-\frac{\eta^2}{G}(d\sigma+F(d\tau+d\sigma))\right)^2\,,    
\end{aligned}
\end{equation}
where
\begin{equation}
\label{eq:ds3}
ds^2_3=G\,\frac{d\rho^2}{\rho^2+1}-\eta^2(\rho^2+\eta^2) \,d\tau^2+\eta^2\rho^2\,d\sigma^2+\eta^2\rho^2 F\,(d\tau+d\sigma)^2\,,    
\end{equation}
and
\begin{subequations}
\begin{equation}
 G=1-\frac{1-\eta^2}{\rho^2+1}\left(\frac{\rho^2}{\rho^2+1}\right)^{\!\tilde{n}}\quad,\quad F =\frac{1-\eta^2}{\eta^2}\left[1-\left(\frac{\rho^2}{\rho^2+1}\right)^{\!\tilde{n}} \right]\,,   
\end{equation}
\begin{equation}
 \Lambda=\left[1-\frac{1-\eta^2}{\rho^2+1}\left(\frac{\rho^2}{\rho^2+1}\right)^{\!\tilde{n}} \sin^2\theta\right]^\frac{1}{2}\,. 
\end{equation}
\end{subequations}
Note that we have set the AdS$_3$ radius to one. The 3D metric $ds^2_3$ is the Einstein metric in the asymptotically AdS$_3$ space: a crucial simplification, which relies on our particular choice\footnote{Replacing $s^{(f)}_1$ with some $s^{(f)}_k$ for $k>1$ spoils this feature of the metric.} for $O_H$, consists in the fact that $ds^2_3$ does not depend on the $S^3$ angle $\theta$; in essence this is what guarantees that the wave equation describing small perturbations around this metric factorises and reduces to an ODE in the radial variable $\rho$.

We can now highlight some of the relevant features of this geometry. For generic values of $\eta$ $ds^2_6$ tends to AdS$_3\times S^3$ asymptotically, i.e. for $\rho\to\infty$. The geometry caps off smoothly for $\rho\to 0$ and it is everywhere regular and horizon-less. For $\eta=1$, the metric is global AdS$_3\times S^3$ for all $\rho$. The limit $\eta\to 0$ is more interesting: if $\tilde{n}>0$ and $\eta\ll 1$ the geometry develops a long AdS$_2$ throat, where the radius of the circle parametrised by $\sigma$ stabilises to a constant finite value, for $1\ll \rho\ll \frac{\sqrt{\tilde{n}}}{\eta}$; as far as $\eta$ is different than zero, the throat ends into a smooth cap. By an appropriate scaling of the coordinates, one can show that in the limiting case $\eta\to 0$ the geometry reduces to that of the Strominger-Vafa black hole in the near-horizon limit: Setting
\begin{equation}
\rho=\frac{r}{a}\quad,\quad \eta=\frac{a}{a_0}
\end{equation}
and sending $a\to 0$ keeping $r$ and $a_0$ fixed, yields $\Lambda\to 1$, $G\to 1$, $F\to \frac{\Tilde{n}a_0^2}{r^2}$ and so
\begin{equation}
ds^2_6 = \frac{dr^2}{r^2}+\frac{r^2}{a_0^2}(-d\tau^2+d\sigma^2)+\tilde{n}(d\tau+d\sigma)^2 + d\theta^2+\sin^2\theta\left(d\varphi_1+d\tau\right)^2+\cos^2\theta(d\varphi_2+d\sigma)^2\,,
\end{equation}
which is the direct product of the extremal BTZ geometry times $S^3$, after the coordinate shifts $\varphi_1\to \varphi_1-\tau$, $\varphi_2\to \varphi_2-\sigma$, which implement the spectral flow from the NS to the R sector. To bring $ds^2_3$ into the standard BTZ form
\begin{equation}
ds^2_3=\frac{\hat r^2}{(\hat r^2-r_0^2)^2}d\hat r^2-\frac{(\hat r^2-r_0^2)^2}{\hat r^2}d\tau^2+\hat r^2\left(d\sigma+\frac{r_0^2}{\hat r^2}d\tau\right)^2
\label{eq:standardbtz}
\end{equation}
one has to perform the redefinitions
\begin{equation}
\hat r^2=\frac{r^2+n\,a_0^2}{a_0^2}\quad,\quad r_0^2=\tilde{n}\,.
\end{equation}
Note that the properties of the dual metric $ds^2_6$, especially in the small $\eta$ regime, are qualitative different when $\tilde n=0$, and hence the momentum charge vanishes: the two charge geometry with non-vanishing $\eta$ interpolates between an asymptotic AdS$_3\times S^3$ region for large $\rho$ and a smooth cap around $\rho\to 0$, without going through an AdS$_2$ throat; the $\eta\to 0$ limit of the 3D geometry $ds^2_3$ is the massless BTZ black hole, where the regular horizon of the extremal BTZ solution degenerates into a naked singularity. 

In summary, we expect that the state $O_H$, though atypical, captures some feature of typical black hole states when $\tilde n>0$ and $\eta$ is small. In particular the $\eta\to 0$ limit of observables computed in $O_H$ should share some properties with the classical (large $N$) limit of black hole observables.  

\subsection{The wave equation}

To summarise, the gravitational description of the heavy state $O_H$ is given by a smooth, horizon-less and asymptotically AdS$_3\times S^3$ geometry, $ds^2_6$, and the light operator $O_L$ is dual to a massless minimally coupled scalar around the AdS$_3\times S^3$ vacuum. 

To compute the four-point correlator between two $O_H$'s and two $O_L$'s, one needs to study the linearised perturbations of $ds^2_6$ induced by $O_L$: this is in general a complicated problem since the perturbations dual to different light states, $O_L$ and $O'_L$, are generically coupled already at linear order; this happens whenever there is a non-vanishing three-point coupling $\langle O_H O_L O'_L\rangle$. The $SO(N_f)$ symmetry of supergravity guarantees that the latter three-point couplings are trivial when $O_H$ and $O_L$ carry different flavour indices. For heavy states of the type considered here, where the geometry of the compact space $\mathcal{M}$ is trivial, an example of a light state belonging to a different tensor multiplet is a perturbation of the $\mathcal{M}$ metric: it was indeed proved in \cite{Bombini:2017sge} that such perturbations satisfy the massless d'Alambert equation even in the non-trivial $ds^2_6$ background. We will make this simplifying choice for the heavy and light operators in the following, so that the field $\Phi_L$ dual to $O_L$ satisfies the wave equation 
\begin{equation}\label{6Dwaveeq}
\Box_6 \Phi_L=0
\end{equation}
at liner order around the heavy state metric $ds^2_6$, where $\Box_6$ is the d'Alambert operator associated with this 6D metric. The solution of this wave equation further simplifies thanks to the structure of $ds^2_6$ given in \eqref{6DEinstein}, and in particular to the $\theta$-independence of the 3D metric $ds^2_3$ noted above: it was shown in \cite{Bena:2017upb} that this allows to make a factorised ansatz
\begin{equation}\label{eq:armdec}
\Phi_L = \psi(\rho,\tau,\sigma) Y_{k-1}(\theta,\varphi_1,\varphi_2) \,,
\end{equation}
where $Y_{k-1}$ is a scalar spherical harmonic of order $k-1$ with respect to the {\it round} metric on $S^3$. We can eliminate a further minor computational nuisance by assuming that $Y_{k-1}$ does not depend on $\varphi_1$ and $\varphi_2$, or equivalently by choosing the $SU(2)_L\times SU(2)_R$ descendant of $O_L$ with $j=\bar j=0$, 
which can be done for every odd $k$. Then one can show that the 6D d'Alambert equation \eqref{6Dwaveeq} implies that the 3D field $\psi(\rho,\tau,\sigma)$ satisfies
\begin{equation}
\left[\Box_3 -\frac{\Delta(\Delta-2)}{G}\right] \psi(\rho,\tau,\sigma)=0\,,
\end{equation}
where $\Box_3$ is the d'Alambertian associated with the 3D Einstein metric $ds^2_3$ \eqref{eq:ds3} and 
\begin{equation}
\Delta=h+\bar h=k+1
\end{equation}
is the total dimension of $O_L$. Note that the rescaling of the mass term $m^2=\Delta(\Delta-2)$ by the function $G$ originates from the non-triviality of the volume of the asymptotically-$S^3$ part of the 6D metric. One can now exploit the independence of $ds^2_3$ on $\tau$ and $\sigma$ to write
\begin{equation}\label{eq:ftts}
\psi(\rho,\tau,\sigma)=\frac{1}{(2\pi)^2}\sum_{\ell\in\mathbb{Z}}\int \!d\omega\, e^{i\omega \tau +i \ell\sigma} \,g(\omega,\ell)\,\psi(\rho)\,,
\end{equation}
and reduce the problem to an ODE for $\psi(\rho)$:
\begin{equation}
\begin{aligned}
&\psi''(\rho) + \frac{1+3\rho^2}{\rho(1+\rho^2)}\psi'(\rho)-\frac{\Delta(\Delta-2)}{\rho^2+1}\psi(\rho) + \\ &\frac{\rho^2 (\ell-\omega) \left[(\ell-\omega)\left(1 - (1-\eta^2) \left(\frac{\rho^2}{1+\rho^2} \right)^{\!\tilde{n}} \right)-2 \eta^2 \ell\right]-\eta^4 \ell^2}{\eta^4 \rho^2 (1+\rho^2)^2} \psi(\rho) =0\,.
\end{aligned}
\label{eq:gennwave}
\end{equation}

\section{HHLL correlators at strong coupling}
\label{sec:strongcorr}

The aim of this section is to solve~\eqref{eq:gennwave} with the boundary conditions dictated by the AdS/CFT dictionary. This was done analytically in~\cite{Bombini:2017sge} for $\tilde n=0$ and our main goal is to extend this analysis to the simplest non-trivial microstate solution carrying a momentum charge. Since the supergravity solutions describing these microstates are fully regular (see~\cite{Bena:2017xbt} for a discussion focusing on the case of interest here), the standard prescription is to start from a solution that is free of singularities at the centre of space $\rho\to 0$ 
\begin{equation}\label{eq:regularity}
\psi_{reg}(\rho) = \rho^{|\ell|} \left(1+\mathcal{O}\left(\rho\right)\right)  \quad \text{(regularity at $\rho = 0$)} \,.
\end{equation}
The holographic information is encoded in the behaviour of such solution close to the AdS boundary ($\rho\to \infty$). In this region $\psi_{reg}$ will be a superposition of a non normalizable mode scaling as $\rho^{\Delta-d}$ for AdS$_{d+1}$ (source) and a normalizable mode scaling as $\rho^{-\Delta}$ (response). Thus we have
\begin{equation}
\psi_{reg}(\rho) = \mathcal{A}\left(\omega, \ell\right) \rho^{\Delta-2} \left(1+\mathcal{O}\left(\rho^{-2}\right)\right) + \mathcal{B}\left(\omega, \ell\right) \rho^{-\Delta} \left(1+\mathcal{O}\left(\rho^{-2}\right)\right) \,,
\label{eq:boundarybehavior}
\end{equation}
where $\mathcal{A}$ and $\mathcal{B}$ are the so called connection coefficients of the ODE \eqref{eq:gennwave}, since they relate the local solution of a differential equation close to a singularity ($\rho=0$) to the local expansion close to a different singularity ($\rho\to\infty$). The standard approach is to divide~\eqref{eq:boundarybehavior} by $\mathcal{A}$ so as to set the coefficient of the non-normalisable term to $1$. Then, after the Fourier transform~\eqref{eq:ftts}, the non-normalisable contribution is proportional to $\delta(\tau)\,\delta(\sigma)$ and the normalisable term describes the response to such localised perturbation. Thus the quantity we are interested in from the holographic point of view is the ratio of the response to the source which, as discussed below, is the boundary correlator
\begin{equation}\label{eq:bcorr}
G\left(\omega, \ell\right) = \frac{\mathcal{B}\left(\omega, \ell\right)}{\mathcal{A}\left(\omega, \ell\right)}\;, \quad 
\mathcal{C}(u,v) = \mathcal{N} \sum_{\ell \in \mathbb{Z}} \int_{\mathbb{R}} d \omega \, e^{i \omega \tau + i \ell \sigma} G(\omega, \ell) \,, 
\end{equation}
where $\mathcal{N}$ is a constant that depends on the normalization of the operators; $(\tau,\sigma)$ are coordinates on a cylinder with periodicity $\sigma\in[0,2\pi]$ and the relation between $(\tau,\sigma)$ and $(u,v)$ is given in \eqref{planecoordinates}.

According to the standard AdS/CFT dictionary~\cite{Witten:1998qj,Gubser:1998bc}, the gravitational result~\eqref{eq:bcorr} is directly linked to correlators between local operators in the dual conformal theory. In our case the relevant observables are the correlation functions involving two heavy and two light operators which are usually called HHLL correlators. Schematically we have
\begin{equation}\label{eq:corrz}
\mathcal{C}\left(u, v\right) = {|z_c|^{\Delta}} \langle {\mathcal{O}}_H(0) \bar{\mathcal{O}}_H(\infty) \mathcal{O}_L(1) \bar{\mathcal{O}}_L(z_c,\bar{z}_c)\rangle \,,
\end{equation}
where 
\begin{equation}\label{planecoordinates}
  z_c=e^{i (\tau+\sigma)} = e^{i v}\quad,\quad \bar{z}_c = e^{i (\tau-\sigma)} = e^{i u}
\end{equation}
and the factor of $|z_c|^{\Delta}$ is due the transformation $\bar{\mathcal{O}}_L(z_c,\bar{z}_c) = (\partial_u z_c)^h(\partial_v \bar{z}_c)^{\bar h}\bar{\mathcal{O}}_L(u,v)$ from the complex plane, parametrised by $(z_c,\bar{z}_c)$, to the cylinder, with coordinates $(u,v)$. We work with operators normalised to one, so the overall factor of $\mathcal{N}$ in~\eqref{eq:bcorr} is fixed to ensure that~\eqref{eq:corrz} behaves as $|1-z_c|^{-2\Delta}$ in the limit $z_c\to 1$. The operator at infinity should be interpreted with the usual conformal limit $\bar{\mathcal{O}}_H(\infty)=\lim\limits_{z_2\to\infty} z_2^{2h_H} \bar{z}_2^{2\bar{h}_H} \bar{\mathcal{O}}_H(z_2,\bar{z}_2)$. In our case the heavy operators are a linear combination of terms with different conformal dimensions and so the prescription above should be applied to each term of the sum~\eqref{defOH}. When $\tilde n\not=0$ the states in~\eqref{defOH} are not primary even when taken separately for every $p$, as they contain a component which is a Virasoro descendant of the $\tilde n=0$ state plus a genuine primary contribution. This subtlety does not seem to play an important role in our analysis; however it is important that the heavy state~\eqref{defOH} is not a (super)descendant of a chiral primary. On the bulk side this means that the dual solution is not just a change of coordinates of a simpler 2-charge configuration.

The problem of computing the correlator $\mathcal{C}(u,v)$ at strong coupling reduces to the mathematical problem of finding the connection coefficients $\mathcal{A}$ and $\mathcal{B}$ that allow to analytically continue the solution of the differential equation \eqref{eq:gennwave} from one of its singular points, $\rho=0$, where we impose the regularity condition \eqref{eq:regularity}, up to the AdS boundary $\rho\to\infty$, which is another singular point of the equation. In order to characterize the singularities of the ODE it is convenient to perform the change of variables
\begin{equation}
z = \frac{\rho^2}{1+\rho^2} \,, \quad\, \psi(\rho) = z^{-\frac{1}{2}} u(z) \,.
\end{equation}
Then the wave equation will take the Scrh\"{o}dinger form
\begin{equation}
\left(\partial_z^2 + V_n(z)\right) u(z) = 0 \,,
\label{eq:schrodform}
\end{equation}
with
\begin{gather}
    \label{eq:Vgn}
    V_n(z) = \frac{x_0 + x_1 z + x_{\tilde{n}+1} z^{\tilde{n}+1}}{4 \eta^4 z^2 (1-z)}-\frac{\Delta(\Delta-2)}{4z(1-z)^2}\;, \\
    x_0= \eta^4 (1-\ell^2)\,,\quad
    x_1=  \big(\eta^2 (\ell-1) -(\ell-\omega) \big) \big(\eta^2 (\ell+1)-(\ell-\omega) \big) \,, \nonumber \\ 
    x_{\tilde{n}+1} = (\eta^2-1) (\ell-\omega)^2\,.\nonumber
\end{gather}
We call a singularity of the ODE a pole, $z_i$, of the potential $V_n$ of order $q$:
\begin{equation}
    V_n(z) \simeq \frac{C}{(z-z_i)^q} \,, \quad \text{for $z \sim z_i$} \,.
\end{equation}
Since we are dealing with a second order equation, a pole $z_i$ with $q=2$ is a regular singularity; if $q=4$ we say that it is an irregular singularity of rank 1, if $q=3$ an irregular singularity of rank 1/2. As usual, in order to study the point $z=\infty$, one needs to perform the change of variable $z=1/w$, bring the equation to the form~\eqref{eq:schrodform} and finally consider the singularity of the potential at $w=0$.
When $\tilde{n}=0$ the equation has 3 regular singularities respectively at $z=0, 1, \infty$. The corresponding ODE is the hypergeometric equation, and its connection coefficients are well known. When $\tilde{n}=1$ the equation has regular singularities at $z=0, 1$ and an irregular singularity of rank $1/2$ at $z = \infty$ when $x_{\tilde{n}+1}\not=0$. It is the so called reduced confluent Heun equation (see e.g. \cite{ronveaux1995heun}), and its connection coefficients have been recently computed exploiting methods coming from 2d Liouville CFT and its 4d AGT dual $\mathcal{N}=2$ supersymmetric gauge theory \cite{Bonelli:2022ten} as we will review later. When $\tilde{n}=2$ the equation is the so called confluent Heun equation: it has again regular singularities at $z=0,1$, but now the irregular singularity at infinity has rank $1$. However it turns out that for $\tilde{n}=2$ the values of the parameters in~\eqref{eq:Vgn} take a non-generic form that makes the problem more complicated as we will briefly explain later (see footnote \eqref{ft:n=2}). For $\tilde{n}>2$ the equation is not of the Heun type and so cannot be studied with the techniques used in this paper. In the following we will review the case $\tilde{n}=0$ and discuss in detail the case $\tilde{n}=1$.

\subsection{The correlator in the two-charge state: $\tilde{n} = 0$}
When $\tilde n=0$ the potential in \eqref{eq:schrodform} reads
\begin{equation}
V_0 (z) = \frac{\frac{1}{4}-a_0^2}{z^2}+\frac{\frac{1}{4}-a_1^2}{(1-z)^2}+\frac{a_1^2+a_0^2-a_\infty^2-\frac{1}{4}}{z(z-1)} \,,
\end{equation}
where $a_0, a_1, a_\infty$ are the local monodromies near the singularities at, respectively, $z=0, 1, \infty$, and are given by
\begin{equation}
a_0 = \frac{|\ell|}{2} \,, \,\,\,\, a_1 = \frac{\Delta-1}{2}\,, \,\,\,\, a_\infty = \frac{\sqrt{\ell^2(\eta^2-1)+\omega^2}}{2 \eta} \,.
\label{eq:hypdict}
\end{equation}
As anticipated the wave equation is solved by the hypergeometric function, and in particular the regular solution at $z=0$ is
\begin{equation}
    u_{reg}(z) = z^{\frac{1+a_0}{2}} (1-z)^{\frac{1+a_1}{2}} {}_2F_1 \left( \frac{1}{2}+a_0+a_1+a_\infty, \frac{1}{2}+a_0+a_1-a_\infty, 1+2a_0,  z \right) \,.
\end{equation}
Expanding close to the AdS boundary we find
\begin{align}
u_{reg}(z) &= \frac{\Gamma\left(-2a_1\right)\Gamma\left(1+2a_0\right)}{\Gamma\left(\frac{1}{2}+a_0-a_1+a_\infty\right)\Gamma\left(\frac{1}{2}+a_0-a_1-a_\infty\right)} u_+(z) + \nonumber \\ &+\frac{\Gamma\left(2a_1\right)\Gamma\left(1+2a_0\right)}{\Gamma\left(\frac{1}{2}+a_0+a_1+a_\infty\right)\Gamma\left(\frac{1}{2}+a_0+a_1-a_\infty\right)} u_- (z) \,,  \\ \nonumber
u_{\pm}(z) &= z^{\frac{1+a_0}{2}} (1-z)^{\frac{1\pm a_1}{2}} {}_2F_1 \left( \frac{1}{2}+a_0 \pm a_1+a_\infty, \frac{1}{2}+a_0\pm a_1-a_\infty, 1\pm 2a_1, 1-z \right) \,.
\end{align}
Comparing with \eqref{eq:boundarybehavior} we find for the correlator in momentum space
\begin{equation}
G(\omega, \ell) = \frac{\Gamma\left(-2a_1\right)}{\Gamma\left(2 a_1\right)} \frac{\Gamma\left(\frac{1}{2}+a_0+a_1+ a_\infty\right)\Gamma\left(\frac{1}{2}+ a_0+a_1- a_\infty\right)}{\Gamma\left(\frac{1}{2}+a_0- a_1+ a_\infty\right)\Gamma\left(\frac{1}{2}+a_0-a_1-a_\infty\right)} \,.
\label{eq:Ghyp}
\end{equation}

\subsection{The correlator in the three-charge state: $\tilde{n} = 1$}
\label{sec:ntilde1exact}
In the $\tilde{n}=1$ case the wave equation reduces to the reduced confluent Heun, and the potential in \eqref{eq:schrodform} reads
\begin{equation}
V_1(z) = \frac{u-\frac{1}{2}+a_1^2+a_0^2}{z(z-1)} + \frac{\frac{1}{4}-a_1^2}{(1-z)^2}+\frac{\frac{1}{4}-a_0^2}{z^2} - \frac{L^2}{4z} \,,
\label{eq:RCHE}
\end{equation}
where
\begin{equation}
\begin{aligned}
&a_0 = \frac{|\ell|}{2} \,, \quad a_1 = \frac{\Delta-1}{2} \,,
\quad L = \frac{i(\ell-\omega)\sqrt{1-\eta^2}}{\eta^2} =\frac{2i w \sqrt{1-\eta^2}}{\eta^2} \,, \\ &u = \frac{\ell^2(1-\eta^2)+\eta^2- \omega^2}{4\eta^2}=\frac{(p+w)^2(1-\eta^2)+\eta^2-(p-w)^2}{4\eta^2}\,.
\end{aligned}
\label{eq:dict}
\end{equation}
Here $w, p$ are the lightcone momenta defined as
\begin{equation}
p = \frac{\ell+\omega}{2} \,, \,\,\,\, w = \frac{\ell-\omega}{2} \,.
\label{eq:lightconem}
\end{equation}
The reduced confluent Heun equation has regular singularities at $z = 0, 1$ ($\rho = 0, \infty$) with local monodromies parametrized respectively by $a_0, a_1$, and an irregular singularity of rank $1/2$ at $z = \infty$. $u$ is the so called accessory parameter, and $L$ is the modulus of the irregular singularity. When $L=0$ the irregular singularity becomes regular and the equation reduces to the hypergeometric equation. For this ODE the connection coefficients $\mathcal{A}, \mathcal{B}$ as in \eqref{eq:boundarybehavior} have been recently computed explicitly exploiting methods coming from 2d Liouville CFT and its 4d AGT dual $\mathcal{N}=2$ supersymmetric gauge theory \cite{Bonelli:2022ten}. 

Let us review the basic idea behind this method. Details will be discussed in the appendix \ref{app:conn}. Heun equations correspond to the semiclassical limit of BPZ equations \cite{Belavin:1984vu} satisfied by Liouville correlation functions with a degenerate operator \cite{Jeong:2018qpc,Piatek:2017fyn, Maruyoshi:2010iu,Alday:2009fs,Drukker:2009id,Ito:2017iba,Bonelli:2011na}. Regular singularities correspond to primary insertions, while irregular singularities correspond to the insertion of \textit{irregular vertices} obtained by taking collision limits of primary vertices \cite{Gaiotto:2009ma,Gaiotto:2012sf,Marshakov:2009gn,Bonelli:2011aa}. Crossing symmetry of CFT correlators relates the series expansions of the solution of the BPZ equation close to the different singularities, and therefore the connection coefficients of the differential equation solve the crossing symmetry constraints.  Since three point functions of Liouville CFT are known exactly \cite{Dorn:1994xn, Zamolodchikov:1995aa}, the Heun connection coefficients can be read off from the crossing relations. They will be given in terms of the three point functions and semiclassical conformal blocks. 

According to the AGT duality everything can be rephrased in the language of $\mathcal{N}=2$ supersymmetric gauge theories, where conformal blocks correspond to gauge theoretical (instanton) partition functions \cite{Alday:2009aq,Nekrasov:2009rc,LeFloch:2020uop}. Such partition functions can be computed explicitly in terms of combinatorial objects exploiting localization of the path integral every time the conformal block expansion correspond to a weakly coupled expansion in the gauge theory. The resulting functions are the so called Nekrasov function. After taking the semiclassical limit of the CFT, the Nekrasov functions reduce to the so called Nekrasov-Shatashvili (NS) functions. When the conformal block expansion does not correspond to a weakly coupled expansion in the gauge theory, the connection coefficients will be more complicated as we will see later.

Using this method, $\mathcal{A}$ and $\mathcal{B}$ can be computed as a series in either\footnote{In \cite{Bonelli:2022ten} the connection coefficients for the reduced confluent Heun are computed only in the small $L$ regime. In appendix \ref{app:conn} we compute the large $L$ connection coefficients as well.} $L$ or $1/L$. From the point of view of the gauge theory, small $L$ correspond to the weakly coupled expansion of an $SU(2)$ $\mathcal{N}=2$ gauge theory with $N_f = 2$ hypermultiplets in the fundamental representation. On the other hand large $L$ corresponds to a strong coupling expansion of the same theory. From the point of view of the geometry, small $L$ corresponds to $\eta^2 \sim 1$ and large $L$ to $\eta^2 \sim 0$. 

Let us start with $\eta^2 \sim 1$. In this regime the  geometry can be understood as a perturbation of global $AdS_3 \times S^3$, and accordingly the wave equation as a perturbation of the hypergeometric equation. This regime has been already explored in \cite{Giusto:2020mup}. Here $G$ reads (see appendix \ref{app:conn} for more details) 
\begin{equation}
G(\omega, \ell) = \frac{\Gamma\left(-2a_1\right) \Gamma\left(\frac{1}{2} + a_0+a_1+a\right)\Gamma\left(\frac{1}{2} + a_0+a_1-a\right)}{\Gamma\left(2a_1\right)\Gamma\left(\frac{1}{2} + a_0-a_1+a\right)\Gamma\left(\frac{1}{2} + a_0-a_1-a\right)} e^{- \partial_{a_1} F} \,.
\label{eq:smallLG}
\end{equation}
$F$ is the so called NS prepotential. It is given as a convergent series in $L^2 \propto 1-\eta^2$ with coefficients that depend on $a_0, a_1, a$. The parameter $a$ can be understood as the monodromy of a cycle that laces both the singularities at $z = 0, 1$. It is related to the parameters appearing in \eqref{eq:RCHE} via the so called Matone relation \cite{Matone:1995rx,Flume:2004rp}, namely
\begin{equation}
u = \frac{1}{4} - a^2 + \frac{1}{2} L \partial_L F \left(a_0, a_1, a\right) \,.
\label{eq:MatonesmallL}
\end{equation}
The previous expression gives the accessory parameter $u$ as a series in $L^2$. Since in \eqref{eq:smallLG} $a$ appears, in order to evaluate $G$ in terms of gravity parameters through \eqref{eq:dict} we need to invert the Matone relation perturbatively in $L^2$ to obtain $a$ as a function of $u$. We have (see \eqref{eq:Matonea})
\begin{equation}\label{eq:adict}
a=-\frac{\omega}{2}+(1-\eta^2)\frac{2(\ell^2-\omega^2)+\frac{(\ell-\omega)^2\left(\ell^2-\omega^2-\Delta(\Delta-2)\right)}{\omega^2-1}}{8\omega}+\mathcal{O}((1-\eta^2)^2) \,.
\end{equation}
What makes this expression for $G(\omega, \ell)$ convenient is that $F$ admits a very concrete combinatorial expression (see equation \eqref{eq:Fcomb}) that can be algorithmically computed to high orders. Note that equation \eqref{eq:smallLG} has formally the same structure of equation \eqref{eq:Ghyp}. The only difference is that now $a$ and $F$ are series in $L^2$. We can say that equation \eqref{eq:smallLG} resums in $a$ and $F$ all the corrections of the reduced confluent Heun equation close to the hypergeometric point $L=0$. Furthermore, the series expressing $a$ and $F$ in terms of the gravitational parameters is believed to be convergent, making equation \eqref{eq:smallLG} in a sense exact.

We now move to the large $L$ regime. $L$ can be made large by either considering $\eta \sim 0$ or large $w$ with finite $\eta$. However, when $w \gg 1$, $a_0$ in \eqref{eq:dict} diverges as $L$, and thus a series in $L^{-1}$ with coefficients depending on the various $a_i$'s will be ill defined\footnote{\label{ft:n=2}This is similar to what happens in the case $\tilde{n}=2$: here the equation is the confluent Heun equation, and the connection coefficients can be still computed as a series in small $L$ ($\eta^2 \sim 1$) or large $L$ ($\eta^2 \sim 0$). However the dictionary will fix some $a_i$ to be basically proportional to $L$ for any $\omega, \ell$, making the large $L$ series ill defined.}. There are different ways to go around this problem: one can either consider $\omega \to \infty$ with $\ell = 0$, or $w \to \infty$ plus $\eta \to 0$. The latter case is equivalent to just considering $\eta \to 0$ with finite $w$. We will consider this case in the following. 

The solution for the large $L$ connection coefficients looks quite different from the one resulting in \eqref{eq:smallLG}. In fact, from the point of view of Liouville CFT, large $L$ is a different conformal blocks expansion with respect to the one of small $L$. In particular, the two expansions will involve different three point functions and give different-looking results for $G\left(\omega, \ell\right)$. We find (again see appendix \ref{app:conn} for more details) 
\begin{equation}
G(\omega,\ell) = e^{-\partial_{a_1}F_D} \frac{\Gamma\left(-2a_1\right)}{\Gamma\left(2a_1\right)} \frac{\frac{ (4L)^{\frac{g}{2}+2a_0+a_1} e^{L +\partial_g F_D}}{\Gamma\left(\frac{1+g}{2}- a_1\right)} +\frac{(-4L)^{-\frac{g}{2}+2a_0+a_1} 
e^{- L -\partial_g F_D}}{\Gamma\left(\frac{1-g}{2}- a_1\right)}}{\frac{ (4L)^{\frac{g}{2}+2a_0-a_1} e^{L +\partial_g F_D}}{\Gamma\left(\frac{1+g}{2}+ a_1\right)} +\frac{(-4L)^{-\frac{g}{2}+ 2 a_0-a_1} e^{- L -\partial_g F_D}}{\Gamma\left(\frac{1-g}{2}+a_1\right)}}\,.
\label{eq:largeLG}
\end{equation}
$F_D$ is now given as a series in $L^{-1}$: it is still the prepotential of a $\mathcal{N}=2$ theory, but in a strongly coupled phase. In particular $F_D$ does not admit a convenient combinatorial expansion like the one of $F$ and has to be computed order by order using 2d CFT methods (see equation \eqref{eq:FD}). The parameter $g$ plays the role of $a$ in the $\eta^2 \sim 1$ case, and is given by the Matone-like relation
\begin{equation}
u = \frac{g L}{2} + \frac{g^2+1}{8} - \frac{a_1^2}{2}+\frac{1}{4} - a_0^2 + \frac{1}{2} L \partial_L F_D \left(a_0,a_1,g\right) \,.
\label{eq:MatonelargeL}
\end{equation}
Analogously to the previous case this gives $u$ as a series in $L^{-1}$. In order to evaluate $G$ one needs to invert \eqref{eq:MatonelargeL} perturbatively in $L^{-1}$ to express $g = g(u)$ in terms of gravity parameters. From \eqref{eq:FD} we find
\begin{equation}\label{eq:gdict}
g = - i p \left(1+\frac{\eta^2}{2}\right) - i \frac{\eta^2}{8w} \left(p^2+\Delta(\Delta-2)\right) + \mathcal{O}(\eta^4) \,.
\end{equation}
Note that since $F_D$ is even under the reflection $(g,L)\to (-L,-g)$, the solution for $g$ is odd under $(\omega,\ell)\to (-\omega,-\ell)$ and hence the form of the propagator $G$ in \eqref{eq:largeLG} is explicitly even: $G(\omega,\ell)=G(-\omega,-\ell)$, as for the propagators in \eqref{eq:largeLG} and in \eqref{eq:Ghyp}. This follows from the invariance of both the wave equation \eqref{eq:gennwave} and the boundary condition \eqref{eq:regularity} under $(\omega,\ell)\to (-\omega,-\ell)$. We end the section stressing again that, despite the appearances, equations \eqref{eq:smallLG} and \eqref{eq:largeLG} correspond to the same global function expressed in different patches. In particular, if one was able to express $g$ and $F_D$ as a series in $L^2$ instead of $L^{-1}$ one would recover precisely \eqref{eq:smallLG}.

\section{The analytic structure of the correlator in Fourier space}\label{sec:anal-str}

The analysis of the previous section has lead to explicit expressions for the Fourier space correlator $G(\omega,\ell)$ in the two opposite regimes $\eta\to 1$ and $\eta\to 0$. In the following we will extract the physical information contained in $G(\omega,\ell)$ and discuss its analytic structure. Finally, we will show how this structure gets modified in the $\eta\to 0$ limit, when the supergravity geometry dual to the pure heavy state $O_H$ approaches the extremal black hole which describes a thermal ensemble with non-vanishing left-moving temperature in the boundary theory.

Much of the physical information is encoded in the poles of $G(\omega,\ell)$. Their meaning is clarified by thinking at the direct channel ($z_c,\bar z_c\to 0$) OPE decomposition of the HHLL correlator in conformal blocks: 
\begin{equation}\label{eq:OHOLOPE}
\langle {\mathcal{O}}_H(0) \bar{\mathcal{O}}_H(\infty) \mathcal{O}_L(1) \bar{\mathcal{O}}_L(z_c,\bar{z}_c)\rangle = \sum_{\mathcal{O}_{HL}} |C_{HL[HL]}|^2 z_c^{-h_L-h_H}\bar{z}_c^{-\bar{h}_L-\bar{h}_H} \bigg|\mathcal{V}_{[HL]}^{h_H-h_L}(z_c) \bigg|^2 \,,
\end{equation}
where $\mathcal{V}_{[HL]}^{h_H-h_L}(z_c)$ is the Virasoro conformal block normalized such that it starts as $z_c^{h_{[HL]}}(1+\mathcal{O}(z_c))$ as $z_c \to 0$. The sum runs over heavy-light composite operators of (holomorphic) scaling dimension $h_{[HL]}$. A set of states that is always present in the heavy-light OPE have the schematic form
\begin{equation}
\mathcal{O}_{HL} = :\mathcal{O}_H \left(\partial \bar{\partial} \right)^n \partial^{|\ell|} \mathcal{O}_L: \,,
\label{eq:HL}
\end{equation}
{\rm i.e.} they are the non-BPS version of the graviton gas\footnote{When the number of non-mutually BPS supergravitons present is large, it is again possible to derive regular geometries dual to this class of non-BPS graviton gas states~\cite{Ganchev:2021pgs,Ganchev:2021ewa}.}. In the Generalized Free Field (GGF) limit, their dimension is
\begin{equation}\label{genfreedim}
h^{(0)}_{[HL]} = h_H + h_L + \frac{|\ell|}{2} + n \,,
\end{equation}
but in the black hole regime~\eqref{eq:bhregeta} we have a new class of non graviton gas states with the same GFF dimensions~\eqref{genfreedim} and it is natural to expect that they appear in the heavy-light OPE of our correlator. Generically the correlators in momentum space $G(\omega, \ell)$ will have poles on the real $\omega$ axis corresponding to the scaling dimensions of the heavy-light double traces, namely
\begin{equation}
\omega_{n \ell} = h_{[HL]} + \bar{h}_{[HL]} - h_H - \bar{h}_H \,,
\end{equation}
and will be analytic in the rest on the complex plane. When working in the large central charge limit, there will be a large number of operator contributing in the heavy-light channel and the poles should be interpreted as average dimensions over (almost) degenerate states. From the bulk viewpoint these poles can be interpreted as bound state energies of the  Schr\"{o}dinger problem \eqref{eq:schrodform}. In fact $G$ has a poles when $\mathcal{A} = 0$, that is when the coefficient of the non normalizable mode in \eqref{eq:boundarybehavior} vanishes. Since the wavefunction is regular at the origin, these solutions are normalizable. In the eikonal approximation, these poles correspond to orbits of the light probe around the heavy object.

It is useful to consider a dispersive representation for $G$. Let $\tilde{G} = G/\omega^\delta$ be the subtracted correlator where $\delta$ is such that the contour at infinity in the integral appearing in Eq.\eqref{tildeGRes} vanishes. From the Cauchy theorem we find
\begin{equation}\label{tildeGRes}
\tilde{G}(\omega, \ell) = \frac{1}{2\pi i} \int_{\gamma_\omega} \!d\omega'\,\frac{\tilde{G}(\omega', \ell)}{\omega'-\omega} = \sum_n \frac{\text{Res}(\tilde{G},\omega_n)}{\omega-\omega_n} \Rightarrow G(\omega, \ell) = \sum_n \left( \frac{\omega}{\omega_n} \right)^\delta\frac{\text{Res}(G,\omega_n)}{\omega-\omega_n} \,,
\end{equation}
where we have denoted by $\gamma_\omega$ a small contour around $\omega'=\omega$ and we have deformed this contour towards infinity. Note that subtractions can add a pole at $\omega=0$ in $\tilde{G}$. This will add to $G(\omega,\ell)$ a term that grows polynomially in $\omega$, that is a contact term in position space. We will ignore terms of this kind in the following.

We can explicitly write down the Feynman and retarded correlators as
\begin{equation}
\begin{aligned}
&G_F(\omega, \ell) = \sum_{\omega_n>0} \left( \frac{\omega}{\omega_n} \right)^\delta \frac{\text{Res}(G,\omega_n)}{\omega-\omega_n+i\epsilon} + \sum_{\omega_n<0} \left( \frac{\omega}{\omega_n} \right)^\delta \frac{\text{Res}(G,\omega_n)}{\omega-\omega_n-i\epsilon} \,, \\
&G_R(\omega, \ell) = G(\omega+i\epsilon,\ell) = \sum_{n} \left( \frac{\omega}{\omega_n} \right)^\delta \frac{\text{Res}(G,\omega_n)}{\omega-\omega_n+i\epsilon} \,.
\end{aligned}
\end{equation}
The Wightman correlator is given by $G_W(\omega,\ell) = -\text{sign} \omega \, \text{Im} G_R(\omega,\ell)$, where we are taking $\omega$ real. We have
\begin{equation}
\text{Im} G_R(\omega,\ell) = \frac{1}{2i}\left(G(\omega+i\epsilon,\ell) - G(\omega-i\epsilon,\ell)\right) = \sum_n \pi \delta(\omega-\omega_n) \text{Res}(G,\omega_n) \,,
\end{equation}
where we used the identity
\begin{equation}
\text{Im} \left( \frac{1}{x+i\epsilon}\right) = \pi \delta (x) \,.
\end{equation}
In the following two subsections we will show, by examining the explicit form of the propagator for $\tilde n=0$ and $\tilde n=1$, that the real poles of $G$ become dense on the real axis in the limit $\eta \to 0$. Here we briefly discuss what happens to the representation of $G$ given in Eq~\eqref{tildeGRes} in the limit of dense poles. As long as we cannot resolve the discrete nature of the spectrum, we can replace the sum over $n$ with an integral:
\begin{equation}
G(\omega,\ell) \simeq \int d\omega_n \, \frac{\rho_\omega(\omega_n,\ell)}{\omega-\omega_n} \,, \quad \rho_\omega(\omega_n,\ell) = \frac{dn}{d\omega_n} \left( \frac{\omega}{\omega_n} \right)^\delta \text{Res}(G,\omega_n) \,;
\label{eq:sumtoint}
\end{equation}
note that because of the subtraction the density $\rho_\omega(\omega_n,\ell)$ depends both on $\omega_n$ and $\omega$. The dependence on $\omega$ is however trivial and captured by the factor $\left( \frac{\omega}{\omega_n} \right)^\delta$.
The above form for $G(\omega,\ell)$ implies that
\begin{equation}\label{eq:discol}
G(\omega+i\epsilon,\ell) - G(\omega-i\epsilon,\ell) \simeq \int d\omega_n \, \rho_\omega(\omega_n,\ell) 2 \pi i \delta(\omega-\omega_n) =  2\pi i \rho_\omega(\omega,\ell) \,,
\end{equation}
therefore, as the real poles of $G(\omega,\ell)$ become dense, the correlator develops a branch cut on the real axis with discontinuity $\rho_\omega(\omega,\ell)$. Indeed, from the previous equation we have 
\begin{equation}
    \pi \rho_\omega(\omega,\ell) = \text{Im} G_R(\omega,\ell)\,,
\label{eq:rhoImG}
\end{equation}
which is then naturally interpreted as the branch cut discontinuity. The appearance of a branch cut on the real axis is related to the fact that a continuum of states is exchanged in the HL OPE channel ($z_c\to 0$) as $\eta \to 0$, that is when the bulk geometry develops an extremal horizon. At leading order in this limit, the correlator will approach the two point function in the black hole background. Crucially, though the exact propagator does not have complex poles, the quasinormal modes of the black hole will appear in the continuum limit as complex poles of the branch discontinuity, in a similar way to what happens for resonances in quantum mechanical problems. We stress that these resonances arise as an effective description of the spectrum, when being treated as a continuum set instead of a discrete one. In fact, the same mechanism describes the finite $\eta$ physics in the limit of large $\omega$ or, equivalently, small energy resolution. When the separation between energy levels (i.e. real poles) goes beyond the energy resolution, the spectrum gets smeared and effectively a branch cut is created even at finite $\eta$. We will see this mechanism at work in the lightcone limit in configuration space in Section~\ref{sec:configuration}. 

As we have already pointed out when describing the nature of the $\eta\to 0$ limit at the end of Section~\ref{sec:heavystate}, we expect a qualitative similarity between the large $c$ limit of the propagator in the black hole background and the small $\eta$ (or large $\omega$) limit in the pure state $O_H$. In particular, the real spectrum forming a continuum and the appearance of an effective branch cut are phenomena that are expected to characterise also the large $c$ approximation to the unitary description of a thermal ensemble.

\subsection{A warmup: $\tilde n=0$}
Let us start by reviewing what happens for the simple case of $\tilde{n}=0$. The solution is given by \eqref{eq:Ghyp} where $a_0, a_1, a_\infty$ are given in \eqref{eq:hypdict}, that is
\begin{equation}
G(\omega,\ell) = \frac{\Gamma\left(1-\Delta\right)}{\Gamma\left(\Delta-1\right)} \frac{\Gamma\left(\frac{\Delta+|\ell|}{2}+\frac{\sqrt{\omega^2-\ell^2(1-\eta^2)}}{2\eta}\right)\Gamma\left(\frac{\Delta+|\ell|}{2}-\frac{\sqrt{\omega^2-\ell^2(1-\eta^2)}}{2\eta}\right)}{\Gamma\left(\frac{2-\Delta+|\ell|}{2}+\frac{\sqrt{\omega^2-\ell^2(1-\eta^2)}}{2\eta}\right)\Gamma\left(\frac{2-\Delta+|\ell|}{2}-\frac{\sqrt{\omega^2-\ell^2(1-\eta^2)}}{2\eta}\right)} \,.
\end{equation}
The expression is analytic in the complex $\omega$ plane except where the Gamma functions in the numerator have poles. This happens for
\begin{equation}\label{eqpoles2charge}
    \frac{1}{2}+a_0-a_1\pm a_\infty = -n \,, \quad n \in \mathbb{Z}_{\ge0} \,,
\end{equation}
that is $\omega=\pm \omega_n$ with
\begin{equation}
 \omega_{n} = \sqrt{\eta^2 \left(|\ell|+2n+\Delta\right)^2+\ell^2 \left(1-\eta^2\right)}\,. 
\end{equation}
The anomalous dimension, $\gamma_{n\ell}$, of the HL double trace is defined by subtracting from the full dimension, $\omega_n$, its GFF value, \eqref{genfreedim}:
\begin{equation}
\gamma_{n\ell} \equiv \omega_{n} - \left(|\ell|+2n+\Delta \right) \,;
\end{equation}
note that $\gamma_{n\ell}$ starts at $\mathcal{O}\left(1-\eta^2\right)$ as it should. In fact, anomalous dimensions are suppressed by a factor $1/c$, but for heavy states this suppressions is compensated by a factor of $h_H \propto N(1-\eta^2)$. As one can easily check, $\gamma_{n\ell} < 0$ as expected from a bound state energy. The residues at the poles are given by
\begin{equation}
\text{Res}(G,\omega_n) = \frac{-2 (|\ell|+2n+\Delta)}{\Gamma(\Delta)\Gamma\left(\Delta-1\right)} \frac{\eta^2}{\omega_n} \frac{\Gamma\left(\Delta+n\right)\Gamma\left(\Delta+n+|\ell|\right)}{\Gamma\left(1+n+|\ell|\right)\Gamma\left(n+1\right)} \,.
\end{equation}
Since $G(\omega,\ell)$ grows as $\omega^{2\Delta-2}$ for large $\omega$ (times oscillating factors), we can choose $\delta = 2\Delta$ in the subtracted correlator. We have (noting that $\frac{1}{\omega-\omega_n}-\frac{1}{\omega+\omega_n} = \frac{2\omega_n}{\omega^2-\omega_n^2}$)
\begin{equation}
G(\omega,\ell) = \sum_{n\ge0} \left(\frac{\omega^2}{\omega_n^2}\right)^{\Delta} \frac{-4 \eta^2 (|\ell|+2n+\Delta)}{\Gamma(\Delta)\Gamma\left(\Delta-1\right)} \frac{\Gamma\left(\Delta+n\right)\Gamma\left(\Delta+n+|\ell|\right)}{\Gamma\left(1+n+|\ell|\right)\Gamma\left(n+1\right)} \frac{1}{\omega^2-\omega_n^2} \,.
\label{eq:twochargedisp}
\end{equation}
While the poles are close to linearly spaced for $\eta^2 \sim 1$, as $\eta^2 \to 0$ and $n\gg 1$ the separation between the poles reads
\begin{equation}
\omega_{n+1}-\omega_n = \frac{2}{\omega_n} (|\ell|+2n+\Delta) \eta^2 + \mathcal{O}(\eta^4) \,,
\end{equation}
with $\omega_n^2 \simeq \ell^2 + (2\eta n)^2$, where we included the correction of order $(n\eta)^2$ so the result is valid also in the regime where both $n$ and $1/\eta$ are large with $n\eta$ finite. In the small $\eta$ limit poles become dense on the real axis. As explained above, when this happens we can approximate the sum in \eqref{eq:twochargedisp} by an integral, and the correlator will develop a branch cut on the real axis. At leading order as $\eta\to 0$ with $\eta n$ fixed we have
\begin{equation}
\frac{dn}{d\omega_n} \simeq \frac{1}{2\eta} \frac{\omega_n}{\sqrt{\omega_n^2-\ell^2}} \,,  \quad n(\omega_n) \simeq \frac{\sqrt{\omega_n^2-\ell^2}}{2\eta} \,, \quad \frac{\Gamma\left(\Delta+n\right)\Gamma\left(\Delta+n+|\ell|\right)}{\Gamma\left(1+n\right)\Gamma\left(1+n+|\ell|\right)} \simeq n^{2\Delta-2} \,,
\end{equation}
and the discontinuity $\rho$ defined in \eqref{eq:sumtoint} reads
\begin{equation}
\rho_\omega(\omega_n,\ell)=\left(\frac{\omega^2}{\omega_n^2}\right)^\Delta \frac{\text{Im} G_R(\omega_n,\ell)}{\pi} \simeq - 
\frac{1}{\Gamma\left(\Delta\right)\Gamma\left(\Delta-1\right)} \left(\frac{\omega^2}{\omega_n^2}\right)^\Delta \left(\frac{\omega_n^2-\ell^2}{4\eta^2}\right)^{\Delta-1} \,.
\end{equation}
Note that this is just the leading order result in the small $\eta$ and large $n$ expansion with fixed $n\eta$, but all higher order corrections can be resumed by keeping the exact $\eta$-dependence in $a_\infty(\omega_n) = a_\infty|_{\omega=\omega_n}$. In fact, keeping power like corrections, we find from the equation for the poles \eqref{eqpoles2charge} and the definition of $a_\infty$ \eqref{eq:hypdict}:
\begin{equation}
 \frac{dn}{d \omega_n} = \frac{\omega_n}{4\eta^2}\frac{1}{a_\infty(\omega_n)}\,, \quad n(\omega_n) = a_\infty(\omega_n) - \frac{|\ell|+\Delta}{2}\,,
\end{equation}
and thus 
\begin{equation}\label{eq:gr2cf}
\begin{aligned}
\text{Im} G_R(\omega,\ell) &= - 
\frac{\pi }{\Gamma\left(\Delta\right)\Gamma\left(\Delta-1\right)} \frac{\Gamma\left(a_\infty(\omega)+\frac{\Delta-|\ell|}{2}\right)\Gamma\left(a_\infty(\omega)+\frac{\Delta+|\ell|}{2}\right)}{\Gamma\left(a_\infty(\omega)+\frac{2-\Delta-|\ell|}{2}\right)\Gamma\left(a_\infty(\omega)+\frac{2-\Delta+|\ell|}{2}\right)} = \\ &= -
\frac{\pi}{\Gamma\left(\Delta\right)\Gamma\left(\Delta-1\right)} \left(\frac{\omega^2-\ell^2}{4\eta^2}\right)^{\Delta-1} \left(1-\frac{\Delta(\Delta-1)(\Delta-2)}{3(\omega^2-\ell^2)}\eta^2 + \mathcal{O}(\eta^4)\right) \,.
\end{aligned}
\end{equation}

It is important to note that the cut discontinuity $\rho$, or equivalently $\text{Im} G_R$, does not have poles in this case, as expected from the fact that the classical geometry approached as $\eta^2\to0$ is a ``small black hole" with a degenerate horizon of vanishing area and both left and right temperatures equal to zero.

\subsection{A case with finite left-moving temperature: $\tilde{n}=1$}
We now move to the more interesting but difficult case, $\tilde n=1$, which we recall describes a supersymmetric black hole with a non-vanishing left-moving temperature. Our analysis will be done in different steps. We first consider the $\eta^2 \sim 1$ regime, where the result is given by \eqref{eq:smallLG} and the various parameters are defined in \eqref{eq:dict}, \eqref{eq:adict}. Again, this expression is analytic everywhere in the complex $\omega$ plane except where the Gamma functions in the numerator have poles. This happens for
\begin{equation}
\frac{1}{2} + a_0+a_1\pm a = -n \,,\quad n \in \mathbb{Z}_{\ge0} \,.
\end{equation}
This gives two sets of real poles, where, since $a_0$ is proportional to the modulus of $\ell$, we need to distinguish the cases $\ell>0$, $\ell<0$:
\begin{align}
  \omega^{\pm}_{n\ell} &= \pm\left(|\ell|+2n+\Delta\right)+\gamma^{(1)\pm}_{n \ell} (1-\eta^2) + \dots \,,  \\ \gamma^{(1)-}_{n\ell<0} &= - \gamma^{(1)+}_{n\ell>0} (-\ell)\,, \gamma^{(1)-}_{n\ell>0} = - \gamma^{(1)+}_{n\ell<0} (-\ell) \,.\\
  \nonumber
\gamma_{n\ell>0}^{(1)+} &= -\frac{(2n+\Delta)}{2(1+\ell+2n+\Delta)(\ell+2n+\Delta)(-1+\ell+2n+\Delta)} \Bigl(2\ell^3+5\ell^2(2n+\Delta)+ \\ &+\ell(-2+5(2n+\Delta)^2) +(2n+\Delta)\left(6n^2+6n\Delta+(\Delta-1)(2\Delta+1)\right)\Bigr)\,, \label{eq:realpolesh1}
\end{align}
\begin{align}\gamma_{n\ell<0}^{(1)+} &= -\frac{(2\ell-2n-\Delta)}{2(-1+\ell-2n-\Delta)(\ell-2n-\Delta)(1+\ell-2n-\Delta)}\Bigl(12n^3+2n^2(9\Delta-8\ell) \nonumber \\ &+(3\ell-2\Delta-1)(1+\ell-\Delta)\Delta+2n(-1+3\ell^2-8\ell \Delta+\Delta(5\Delta-1))\Bigr) \,, 
\end{align}
By following~\cite{Kulaxizi:2018dxo,Karlsson:2019qfi,Kulaxizi:2019tkd}, one can derive these anomalous dimensions in a particular limit from the phase shift of a light-like geodesic in the geometry dual to the heavy state. At this purpose one should set $n=\bar{m}$ and $\ell = m-\bar{m}$ in $\gamma^{(1)+}_{n \ell>0}$ and send $m, \bar{m}$ to infinity. For the particular heavy state considered here (see~\eqref{defOH} and~\eqref{6DEinstein} for the CFT and the bulk definitions respectively), this analysis was done in~\cite{Giusto:2020mup} obtaining
\begin{equation}
\lim_{m,\bar{m} \to \infty}  \gamma_{n \ell>0}^{(1) +} = -2 m \bar{m} \frac{m^2+2m \bar{m} + 3 \bar{m}^2}{(m+\bar{m})^3} \,,
\end{equation}
which agrees with the limit of~\eqref{eq:realpolesh1} (it is not difficult to extend this check to the next order in $(1-\eta^2)$ and we have done so by finding again perfect agreement). As for $\tilde{n}=0$, the separation between the poles for fixed $\ell$ is close to linear for $\eta \sim 1$. 

To see what happens in the ``black-hole regime" of small $\eta$ we should consider the expression in \eqref{eq:largeLG}, with dictionary \eqref{eq:dict} and \eqref{eq:gdict}. Here the solution develops a long $AdS_2$ throat in the region $1 \ll \rho \ll \eta^{-1}$. Expanding in this region the wave equation \eqref{eq:gennwave} for $\tilde{n}=1$ we find that down the throat the wave function behaves as (see appendix \ref{app:BTZ})
\begin{equation}
\begin{aligned}
&\psi(\rho) \sim c_1 \, e^{-i\frac{w}{\eta^2\rho^2}}\rho^{-i \frac{p}{2}} + c_2 \,e^{i\frac{w}{\eta^2\rho^2}}\rho^{i \frac{p}{2}} \,, \\
&c_1 \sim -i \,e^{\frac{2iw}{\eta^2}} \quad , \quad c_2 \sim e^{i\pi(p+w)} e^{-\frac{2iw}{\eta^2}} \,.
\end{aligned}
\end{equation}
This is the same behavior that one finds close to an extremal BTZ horizon. If we analytically continue $w$ in the lower half plane we get, for $\eta\to 0$
\begin{equation}
\psi(\rho) \sim c_1 e^{-i\frac{w}{\eta^2\rho^2}}\rho^{-i \frac{p}{2}} +\mathcal{O}(e^{-i\frac{w}{\eta^2}}) \,.
\label{eq:imwbc}
\end{equation}
Accordingly when $\text{Im} \, w<0$, recalling that $L\simeq 2iw/\eta^2$, \eqref{eq:largeLG} gives at leading order
\begin{equation}\label{eq:Gimw}
G \simeq \frac{\Gamma\left(1-\Delta\right)}{\Gamma\left(\Delta-1\right)} \left(-\frac{8iw}{\eta^2}\right)^{\frac{\Delta-1}{2}} \frac{\Gamma\left(\frac{\Delta+ip}{2}\right)}{\Gamma\left(\frac{2-\Delta+ip}{2}\right)}\left(1+\mathcal{O}(\eta^2,e^{iw/\eta^2})\right) \,.
\end{equation}
Comparing with \eqref{eq:RBTZmomentum}, we see that this reproduces the retarded correlator $G_R$ in an extremal BTZ background. This was expected since the boundary condition \eqref{eq:imwbc}, up to exponentially small corrections, corresponds to a wave that is purely ingoing down the throat. Exponentially small corrections in \eqref{eq:Gimw} become relevant when $p= -i (\Delta+2n)$, and cancel the divergence of the overall Gamma functions. All in all, \eqref{eq:largeLG} has bumps of order $e^{-iw/\eta^2}$ for $\text{Im}w>0$ at $p= -i (\Delta+2n)$ that correspond to the QNMs of the extremal BTZ black hole approached as $\eta\to 0$.

\begin{figure}
\centering
\begin{subfigure}{.45\textwidth}
  \centering
  \includegraphics[width=1\linewidth]{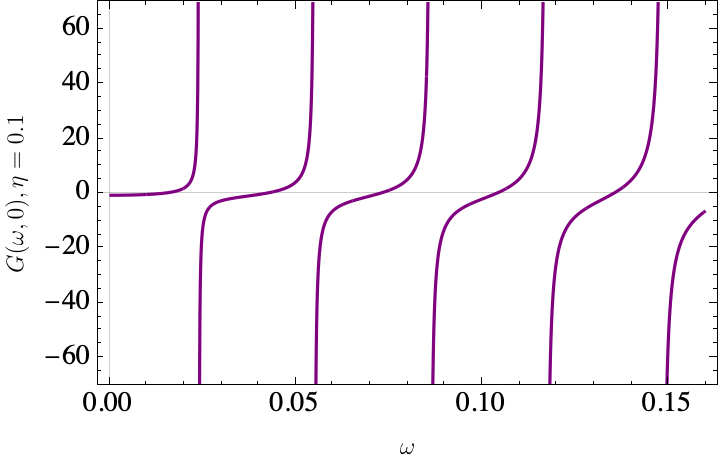}
  \caption{}
  \label{fig:h01a}
\end{subfigure}
\begin{subfigure}{.45\textwidth}
  \centering
  \includegraphics[width=1\linewidth]{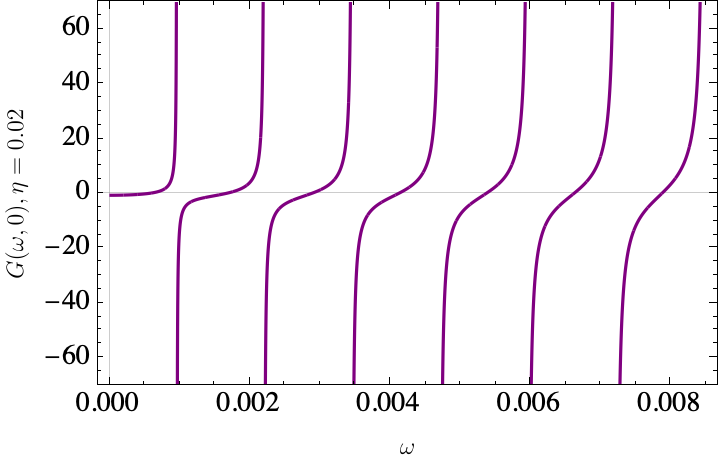}
  \caption{}
  \label{fig:h01b}
\end{subfigure}
\caption{(a) $G(\omega, \ell)$ for $\ell = 0$, $\Delta = 2$ and $\eta = 0.1$. (b) $G(\omega, \ell)$ for $\ell = 0$, $\Delta = 2$ and $\eta = 0.02$. The plots are obtained solving numerical the wave equation with \texttt{NDSolve} in Mathematica imposing regularity at $\rho=0$, and then reading off the value of the correlator at the boundary (see e.g. \cite{Hartnoll:2009sz}).}
\label{fig:h0105}
\end{figure}

Actual poles of \eqref{eq:largeLG} appear when the denominator vanishes, that is for
\begin{equation}
\frac{ (4L)^{\frac{g}{2}+2a_0-a_1} e^{L +\partial_g F_D}}{\Gamma\left(\frac{1+g}{2}+ a_1\right)} +\frac{(-4L)^{-\frac{g}{2}+2a_0-a_1} e^{- L -\partial_g F_D}}{\Gamma\left(\frac{1-g}{2}+a_1\right)} = 0 \,.
\label{eq:zerolargeL1}
\end{equation}
Note that $L$ and $g$ are purely imaginary for $p,w$ real, as can be seen from the dictionary \eqref{eq:dict} and \eqref{eq:gdict}. We can then rewrite \eqref{eq:zerolargeL1} as
\begin{equation}
\begin{aligned}
(-16L^2)^{-g/2} &e^{i\pi\left(-2a_0+a_1\right)\frac{w}{|w|}} e^{-2L-2\partial_g F_D} \frac{\Gamma\left(\frac{1+g}{2}+a_1\right)}{\Gamma\left(\frac{1-g}{2}+a_1\right)} \equiv e^{i \theta(w_n,p)} = -1 \\  &\Rightarrow \theta(w_n,p) = (2n+1)\pi \,,\quad n\in \mathbb{Z}\,.
\label{eq:zerolargeL2}
\end{aligned}
\end{equation}
The ratio of $\Gamma$ functions is a pure phase since $g$ is purely imaginary and to determine the phase $e^{i\pi\left(2a_0+a_1\right)\frac{w}{|w|}}$ we have placed the cut of the function $z^\alpha$ along the negative real $z$ axis. As $\eta \to 0$ we find
\begin{equation}
\theta(w,p) \approx -\frac{4 w}{\eta^2} + 2w + \frac{p}{2} \log \left(\frac{8w}{\eta^2}\right)^2 - i \log \frac{\Gamma\left(\frac{-ip+\Delta}{2}\right)}{\Gamma\left(\frac{ip+\Delta}{2}\right)}+\pi\frac{w}{|w|}\left(-|p+w|+\frac{\Delta-1}{2}\right) \,.
\end{equation}
Since $\theta(w,p)$ contains a term that goes as $w/\eta^2$, in order for \eqref{eq:zerolargeL2} to be satisfied we need $w_n \sim n \eta^2$. However if $w_n \sim \eta^2$ then $L$ in \eqref{eq:dict} becomes of $\mathcal{O}(1)$, and the large $L$ expansion breaks down. To avoid this problem we solve \eqref{eq:zerolargeL2} for $n \sim 1/\eta^2$ so that we still have $L\gg 1$. Then, in the regime $\eta\ll 1$ with $n \eta^2 = \mathcal{O}(1)$, we find
\begin{equation}
\label{eq:realpoleslargeL}
\begin{aligned}
w_n &\approx - (n \eta^2) \frac{\pi}{2}  + \frac{\eta^2}{4}  \Biggl[\frac{p}{2}\log \left(\frac{4 \pi}{\eta^2}(n \eta^2)\right)^2 -\pi((n\eta^2)+1) - i \log \frac{\Gamma\left(\frac{-ip+\Delta}{2}\right)}{\Gamma\left(\frac{ip+\Delta}{2}\right)} \\
&\quad + \pi \frac{w}{|w|}\left(-\left|p-(n\eta^2)\frac{\pi}{2}\right|+\frac{\Delta-1}{2}\right)\Biggr] .
\end{aligned}
\end{equation}
As before, the separation between successive poles is of order $\eta^2$. The same scaling can be observed in the numerical plots for $G$ obtained in figure \ref{fig:h0105} also for $w \ll 1$. Moreover a quite non-trivial check of \eqref{eq:realpoleslargeL} is provided by the WKB computation performed in appendix~\ref{sec:wkb} for the two cases of $\ell=0$ and $p=0$. For $\ell=0$ and $\omega_n=-2w_n=2 p_n$ large and positive, we can use the Stirling approximation
\begin{equation}
- i \log \frac{\Gamma\left(\frac{-ip_n+\Delta}{2}\right)}{\Gamma\left(\frac{ip_n+\Delta}{2}\right)}\approx p_n-\frac{p_n}{2}\log \frac{p_n^2}{4}-\pi \frac{\Delta-1}{2}
\end{equation}
to reduce Eq. \eqref{eq:realpoleslargeL} to
\begin{equation}
\omega_n \approx (n\eta^2) \pi \left[1+\frac{\eta^2}{4}\left(1+\log\frac{\eta^2}{16}\right)\right]+\frac{\pi \eta^2}{2}\Delta\,,
\end{equation}
which agrees with the $\eta\to 0$ limit of \eqref{eq:polesWKBell0}. Similarly for $p=0$ we can take $w_n$ large and negative, which amounts to take $n$ positive in \eqref{eq:realpoleslargeL}, and obtain from \eqref{eq:realpoleslargeL}
\begin{equation}
w_n\approx -\frac{\pi}{2}(n\eta^2)\left[1+\frac{\eta^2}{4}(2+\pi)\right]-\frac{\pi\eta^2}{8}(\Delta+1)\,,
\end{equation}
which agrees with \eqref{eq:polesWKBp0} (after sending $w_n\to -w_n$).

It is interesting to compute the residues of $G$ at the poles $w_n$ found above. We first rewrite equation \eqref{eq:largeLG} as
\begin{equation}
\begin{aligned}
G = &\frac{\Gamma\left(-2a_1\right)}{\Gamma\left(2a_1\right)} (4 L)^{2a_1} e^{-\partial_{a_1}F_D} \frac{\Gamma\left(\frac{1+g}{2}+a_1\right)}{\Gamma\left(\frac{1+g}{2}-a_1\right)} \times \\ &\times \frac{1+(-16L^2)^{-g/2} e^{i\pi\left(-2a_0-a_1\right)\frac{w}{|w|}} e^{-2L-2\partial_g F_D} \frac{\Gamma\left(\frac{1+g}{2}-a_1\right)}{\Gamma\left(\frac{1-g}{2}-a_1\right)}}{1+(-16L^2)^{-g/2} e^{i\pi\left(-2a_0+a_1\right)\frac{w}{|w|}} e^{-2L-2\partial_g F_D} \frac{\Gamma\left(\frac{1+g}{2}+a_1\right)}{\Gamma\left(\frac{1-g}{2}+a_1\right)}} \,.
\end{aligned}
\end{equation}
Then
\begin{equation}
\begin{aligned}
\text{Res} (G,w_n) = & \frac{\Gamma\left(-2a_1\right)}{\Gamma\left(2a_1\right)} (4 L)^{2a_1} e^{-\partial_{a_1}F_D} \frac{\Gamma\left(\frac{1+g}{2}+a_1\right)}{\Gamma\left(\frac{1+g}{2}-a_1\right)}\bigg|_{w=w_n} \times\\
&\times\frac{1+(-16L^2)^{-g/2} e^{i\pi\left(-2a_0-a_1\right)\frac{w}{|w|}} e^{-2L-2\partial_g F_D} \frac{\Gamma\left(\frac{1+g}{2}-a_1\right)}{\Gamma\left(\frac{1-g}{2}-a_1\right)}\bigg|_{w=w_n}}{\partial_w \left[(-16L^2)^{-g/2} e^{i\pi\left(-2a_0+a_1\right)\frac{w}{|w|}} e^{-2L-2\partial_g F_D} \frac{\Gamma\left(\frac{1+g}{2}+a_1\right)}{\Gamma\left(\frac{1-g}{2}+a_1\right)}\right] \bigg|_{w=w_n}} \,.
\end{aligned}
\end{equation}
We have
\begin{equation}
\begin{aligned}
&\frac{(4 L)^{2a_1}\Gamma\left(\frac{1+g}{2}+a_1\right)}{\Gamma\left(\frac{1+g}{2}-a_1\right)} \left(1+ \frac{(-16L^2)^{-g/2} e^{i\pi\left(-2a_0-a_1\right)\frac{w}{|w|}} e^{-2L-2\partial_g F_D}\Gamma\left(\frac{1+g}{2}-a_1\right)}{\Gamma\left(\frac{1-g}{2}-a_1\right)} \right)\bigg|_{w=w_n} = \\ &=  \left((4L)^{2a_1}\frac{\Gamma\left(\frac{1+g}{2}+a_1\right)}{\Gamma\left(\frac{1+g}{2}-a_1\right)} - \text{c.c.} \right) \bigg|_{w=w_n} = 2i \frac{\Gamma\left(2a_1\right)}{\Gamma\left(-2a_1\right)} \text{Im} G_R^{(BTZ)} (-iL, ig) |_{w=w_n} \,,
\end{aligned}
\end{equation}
where we noted, comparing with equation \eqref{eq:imgrbtz}, that this is precisely $\text{Im} G_R$ of a BTZ black hole with lightcone frequencies $(w,p)=(-iL,ig)$, up to an irrelevant normalization factor. Moreover
\begin{equation}
\begin{aligned}
&\partial_w (-16 L^2)^{-g/2} e^{i\pi\left(-2a_0+a_1\right)\frac{w}{|w|}} e^{-2L-2\partial_g F_D} \frac{\Gamma\left(\frac{1+g}{2}+a_1\right)}{\Gamma\left(\frac{1-g}{2}+a_1\right)}\bigg|_{w=w_n} = \partial_w e^{i\theta(w,p)}\bigg|_{w=w_n} = \\ &= \frac{1}{2} \left(\log(-16L^2)+4\partial_g^2 F_D -\sum_\pm \psi\left(\frac{1\pm g}{2}+a_1\right)\right) \partial_w g + \\ &\quad+i\pi+\left(2+\frac{g}{L}+2 \partial_g\partial_L F_D\right) \partial_w L \biggl|_{w=w_n}\,,
\end{aligned}
\end{equation}
where $\theta(w,p)$ is the phase defined in \eqref{eq:zerolargeL2} and we have used that $e^{i\theta(w_n,p)}=-1$.
Finally we get
\begin{equation}
\text{Res}(G,w_n) = 2i e^{-\partial_{a_1}F_D} \frac{\text{Im} G_R^{BTZ}(-iL,ig)|_{w=w_n}}{\partial_w e^{i\theta(w,p)}\bigg|_{w=w_n}} \,.
\label{eq:Reswn}
\end{equation}
Note that $\text{Im} G_R^{BTZ}(-iL,ig)$ has complex poles at
\begin{equation}
1\pm g+2 a_1 = -n \,, \quad n \in \mathbb{Z}_{\ge0} 
\end{equation}
corresponding to the quasinormal modes of the extremal BTZ black hole approached in the $\eta \to 0$ limit, but the full residue is regular at these points since the digamma functions in $\partial_w e^{i\theta(w,p)}$ have poles in the same positions. In fact $G$ has to be regular outside the real axis for any finite $\eta$, being a unitary 4 point function in momentum space. This can be easily checked in the $\eta^2 \sim 1$ limit from the expression \eqref{eq:smallLG}, whose only poles are the ones in \eqref{eq:realpolesh1}. Things are different in the $\eta \to 0$ limit. To elucidate the analytic structure of the propagator in the small $\eta$ limit, we write down a dispersive representation for $G$ analogous to the one in \eqref{tildeGRes} but this time in the variable $w$, which is more natural to use than $\omega$ when $\tilde n\not=0$ because $w$ is the variable that describes the right-moving sector where the temperature vanishes. Proceeding as before and replacing the sum over $n$ by an integral over $w_n$, we write
\begin{equation}
G(w,p) = \sum_{n} \left(\frac{w^2}{w_n^2}\right)^{\Delta} \frac{\text{Res}(G,w_n)}{w-w_n} \simeq \int dw_n \, \frac{\rho_w(w_n,p)}{w-w_n} \,;
\end{equation}
since Eq. \eqref{eq:zerolargeL2} implies
\begin{equation}
\frac{dn}{dw_n} = \frac{1}{2\pi} \frac{d\theta(w_n,p)}{dw_n} = \frac{1}{2\pi i} e^{-i\theta(w_n,p)} \partial_{w_n} e^{i\theta(w_n,p)} = - \frac{1}{2\pi i} \partial_{w_n} e^{i\theta(w_n,p)} \,,
\label{eq:ntow}
\end{equation}
and using the form of the residue given in \eqref{eq:Reswn}, we find that the discontinuity $\rho_w(w_n,p)$ can be written in the simple form
\begin{equation}
\rho_w(w_n,p) = -\frac{1}{\pi} \left(\frac{w^2}{w_n^2}\right)^\Delta e^{-\partial_{a_1}F_D} \text{Im}G_R^{BTZ}(-iL,ig)\bigg|_{w=w_n} \,.
\end{equation}
Finally, we can use \eqref{eq:rhoImG} to deduce the imaginary part of $G_R$:
\begin{equation}
\text{Im}G_R(w,p) = -e^{-\partial_{a_1}F_D} \text{Im} G_R^{BTZ}(-iL,ig) \,.
\end{equation}
 The previous formula relates the black hole, or $\eta\to 0$, limit of the propagator $G_R$ in the pure state $O_H$ to the BTZ  propagator $G_R^{BTZ}$; all power like corrections in $\eta^2$ to $G_R$ are encoded in the functions $F_D$, $g$ and $L$. It is interesting to compare the two cases $\tilde n=0$ and $\tilde n=1$. In both situations the propagator $G$ develops a branch cut when $\eta\to 0$, but while for $\tilde n=0$ the discontinuity across the cut does not have complex poles, for $\tilde n=1$ the cut discontinuity has poles in the complex $(w,p)$-plane corresponding to the BTZ QNMs
 .
\section{Configuration space}
\label{sec:configuration}
So far we focused on the analytic structure of the correlator $G(\omega,\ell)$ in Fourier space. It is also useful to perform the Fourier transform in \eqref{eq:bcorr} and study the correlator $\mathcal{C}(u,v)$ in configuration space. This allows a more direct analysis of the OPE expansion of the correlator and the interpretation of the operators exchanged in the various OPE channels. We will concentrate in particular on the light-cone $u\to 0$ expansion, which contains information on the exchanged operators with the lowest values of the right-moving conformal dimension $\bar h$.

As usual, we have access to either of the two opposite limits, $\eta\to 1$ or $\eta\to 0$. For $\eta \to 1$ and $w$ not too large, such that $L\sim w \sqrt{1-\eta^2}\ll 1$, one can use the form \eqref{eq:smallLG} of the propagator and the small-$L$ expansion of the prepotential $F$. This computation will provide the corrections to the vacuum two-point function $\frac{1}{|1-z_c|^{2\Delta}}$, to which the correlator reduces for $\eta=1$. The first non-trivial correction of order $1-\eta^2$ represents the correlator with four light operators, where the heavy operator $\mathcal{O}_H$ has been replaced by its single particle constituent: $\mathcal{O}_H\to L_{-1}^n \sigma_1^{(f)}$. This LLLL correlator has already been computed in \cite{Bombini:2019vnc} and it can be expressed in terms of the Bloch-Wigner function appearing in the standard Witten diagrams. Similarly, corrections of order in $(1-\eta^2)^p$ can be identified with correlators where $\mathcal{O}_H$ is replaced by the $p$-trace operator $(L_{-1}^n \sigma_1^{(f)})^p$; as in \cite{Ceplak:2021wzz}, these correlators are expected to be expressible in terms of Bloch-Wigner-Ramakrishnan polylogarithms, which generalize the Bloch-Wigner function. Though interesting in themselves, these results are not relevant for the black hole regime and we will not discuss the $\eta\to 1$ limit further in this article.   

We now switch to the small $\eta$ limit: more precisely we perform a double expansion in both $\eta\to 0$ and $w^{-1}\to 0$, where we can use the perturbative expression of $F_D$ for large $L\approx \frac{2 i w}{\eta^2}$. This approximation will provide the $u\to 0$ light-cone expansion of the correlator, where each coefficient is also expanded in $\eta$. To simplify the computation we will take the decompactification limit in which the radius of the CFT spatial circle is taken to infinity ($\sigma\in (-\infty,\infty)$); this replaces the sum over $\ell$ in \eqref{eq:bcorr} by an integral. We concentrate on the Feynman propagator $G_F$ and thus compute
\begin{equation}\label{anti-Fourier}
\begin{aligned}
\mathcal{C}(u,v) &= \mathcal{N} \int_{\mathbb{R}^2} d w \,d p\, e^{i p v - i w u} \,G_F(w, p) = \\ &= \mathcal{N} \int_{\mathbb{R}^2} d w \,d p\, e^{i p v - i w u} \, \sum_n \left(\frac{w^2}{w_n^2}\right)^\Delta \frac{\text{Res}(G(w,p),w_n)}{w-w_n \pm i \epsilon} \,,
\end{aligned}
\end{equation}
where the $\pm$ sign is $+$ whenever  $w_n>0$ and $-$ for $w_n<0$ according to the Feynman prescription. We can first exchange the $n$-sum with the $w$-integral and easily evaluate this last integral; without loss of generality we assume $u>0$, close the contour on the lower half plane and pick the poles with $w_n>0$; this yields\footnote{We will conveniently reabsorb in $\mathcal{N}$ irrelevant normalisation factors that arise at various steps in the computation below.}:
\begin{equation}
\mathcal{C}(u,v) = \mathcal{N} \sum_n \int d p\, e^{i p v - i w_n u} \,\text{Res}(G(w,p),w_n) \,.
\end{equation}
We have seen in section~\ref{sec:ntilde1exact} that after taking the $L\to \infty$ limit the Fourier-space propagator $G(w,p)$ develops a cut, and the sum over $n$ can be approximated by an integral over $w_n$, which in the following we will rename as $w$; using \eqref{eq:Reswn} and \eqref{eq:ntow} one finds: 
\begin{equation}
\mathcal{C}(u,v) = \mathcal{N} \int_0^\infty dw \int_\mathbb{R} dp \, e^{ipv-iw u} e^{-\partial_{a_1}F_D} \text{Im} G_R^{BTZ}(-iL,ig) \,.
\end{equation}
 We report here the expression of $\text{Im} G_R^{BTZ}(-iL,ig)$ for convenience:
\begin{equation}
\text{Im} G_R^{BTZ}(-iL,ig) =  \frac{\Gamma\left(-2a_1\right)}{\Gamma\left(2a_1\right)} \left((iL)^{2a_1}\frac{\Gamma\left(\frac{1+g}{2}+a_1\right)}{\Gamma\left(\frac{1+g}{2}-a_1\right)} - (-iL)^{2a_1}\frac{\Gamma\left(\frac{1-g}{2}+a_1\right)}{\Gamma\left(\frac{1-g}{2}-a_1\right)} \right) \,.
\end{equation}
Since $g = -ip + \mathcal{O}(\eta^2)$, when $v>0$ the contour should be closed on the upper half plane, and poles of $\Gamma\left(\frac{1-g}{2}+a_1\right)=\Gamma\left(\frac{\Delta-g}{2}\right)$ will contribute. Vice versa, when $v<0$ the contour should be closed on the lower $p$ and the relevant poles will be the ones of $\Gamma\left(\frac{1+g}{2}+a_1\right)$. 

We will assume $v>0$ in the computation that follows. In the variable $g$, the relevant poles are simply given by 
\begin{equation}
g_n=\Delta+2n\quad,\quad n = 0,1,\dots\,.
\end{equation}
The residue of $e^{-\partial_{a_1}F_D} \text{Im} G_R^{BTZ}(-iL,ig) e^{i p v}$ at $g_n$ is
\begin{equation}\label{eq:resn}
\mathrm{Res}_{g_n}= \frac{(-1)^n \,e^{- \partial_{a_1} F_D|_{g=g_n}}}{n! \Gamma(1-\Delta-n)} \frac{\Gamma\left(1-\Delta\right)}{\Gamma\left(\Delta-1\right)}\,\left(- \frac{2w\sqrt{1-\eta^2}}{\eta^2}\right)^{\Delta-1}  \,\left(\frac{dg}{dp}\Big|_{g=g_n}\right)^{-1} e^{i p(g_n) v}\,.
\end{equation}
The relation between $g$ and $p$ needed to compute $\frac{dg}{dp}$ and $p(g)$ is given by \eqref{eq:MatonelargeL}, where $F_D$ is only known perturbatively in $L^{-1}$; note that the coefficients of $L^{-n}$ in this expansion are growing functions of $a_0\propto p+w$ and thus to compute the correlator at a given order in $w^{-1}$ for large $w$ (or equivalently at a given order in the $u\to 0$ light-cone limit) requires re-summing an infinite number of terms in the $L^{-1}$ series. To suppress the higher order terms in $L^{-1}$ we take a further $\eta\to 0$ limit. We keep terms up to order $L^{-2}$ in $F_D$, and thus we can compute the correlator up to corrections of order $w^{-2}$ in the $w^{-1}$ expansion and $\eta^4$ in the $\eta$ expansion. The $w$-dependence of the generic term in this approximation is of the form $w^{\Delta-1-a}$, with $a=0,1,2$. Thus we have to compute integrals of the form 
\begin{equation}\label{eq:wint}
\int_0^\infty dw\,w^{\Delta-1-a}\,e^{-i w u}=\Gamma(\Delta-a)\,(i u)^{-\Delta+a}\,,
\end{equation}
where to make the integral converge one should analytically continue the variable $u$ appropriately.

We first compute the corrections with $a=0,1$ for generic $\Delta$: this is done by using the perturbative form for $F_D$ in \eqref{eq:FD}, performing the sum of the residues \eqref{eq:resn} over $n=0,1,\ldots,\infty$, using \eqref{eq:wint} to perform the final $w$-integral and fixing the normalization $\mathcal{N}$ so that $\mathcal{C}(u,v)\to (u\,v)^{-\Delta}$ for $u,v\to 0$. We obtain
\begin{equation}\label{eq:conf01}
\begin{aligned}
\mathcal{C}(u,v)&=\left[\frac{1-\frac{\eta^2}{2}-\frac{\eta^4}{4}}{\sinh \left(\Big(1-\frac{\eta^2}{2}-\frac{\eta^4}{4}\Big)v\right)\,u}+O(\eta^6)\right]^\Delta \times \\ \times &\Bigg[1+\frac{u\,\eta^2}{32\,\sinh^2 v} \frac{\Delta}{\Delta-1}\Bigg(-(\Delta+1)(2+5\eta^2)\sinh{2v}+ \\ +v\,\Big(2(2+\eta^2) \Delta+& 6\eta^2+4(1+\eta^2\Delta)\cosh{2v}\Big)+4\,\eta^2 v^2 \,(\Delta+1)\coth{v}\Bigg)+O(u^2,\eta^6)\Bigg]\,.
\end{aligned}
\end{equation}

Before commenting on the meaning of the result above, we note that the case $a=\Delta$ in \eqref{eq:wint} is special and can be defined as a limit
\begin{equation}
\int dw\,w^{-1}\,e^{-i w u}\approx -\log u\,,
\end{equation}
up to $u$-independent terms. From the CFT point of view, the logarithmic terms in the light-cone $u\to 0$ OPE are associated with the exchange of non-protected operators whose dimensions receive anomalous corrections in the large $c$ expansion. These terms thus encode interesting dynamical information on the strong coupling regime of the CFT. Since our approximation for $F_D$ allow us to compute terms with $a\le 2$ and in the D1-D5 CFT one should have $\Delta\ge 2$, we have access to logarithmic terms only for $\Delta=2$. Note also that at this order in the $w^{-1}$ expansion a new class of corrections are generated from the terms proportional to $L^{-\frac{g}{2}}$ in \eqref{eq:largeLG} -- these terms are indeed suppressed by $L^{-g}\sim w^{-\Delta}$ with respect to the leading term. One can however show that these type of corrections give rise to terms that are analytic as $u\to 0$, and can thus be consistently isolated from the $\log u$ terms discussed above. The logarithmic part of the correlator for $\Delta=2$ and up to $O(\eta^4)$ is
\begin{equation}\label{eq:conflog}
\begin{aligned}
\mathcal{C}(u,v)|_{\log}^{\Delta=2}&=\frac{\eta^4\,\log u}{32\sinh^2 v}\left[\frac{67+35\cosh{2v}}{\sinh^2{v}}-v\frac{(59+13\cosh{2v})\cosh{v}}{\sinh^3 v}+ \right.\\&\left.-v^2\frac{33+26\cosh{2v}+\cosh{4v}}{2\sinh^4 v}\right]\,.
\end{aligned}
\end{equation}

\subsection{Block decomposition and CFT interpretation}

The results \eqref{eq:conf01} and \eqref{eq:conflog} are not particularly enlightening but their CFT interpretation becomes much more transparent after expressing them in terms of Virasoro blocks, which resum the contributions of all the Virasoro descendants of a particular primary. Since our results follow from classical supergravity, we need the large $c$ HHLL blocks which were derived in \cite{Fitzpatrick:2014vua,Fitzpatrick:2015zha}. To write these blocks, it is convenient to introduce the ``reduced" dimension, $h_H^{[0]}$, of the heavy external states
\begin{equation}
h_H^{[0]} \equiv h_H-\frac{j_H^2}{N}=N(1-\eta^2)\left(\tilde{n}+\frac{1+\eta^2}{4}\right)\,,
\end{equation}
where one subtracts from the total dimension the $SU(2)$ Sugawara contribution and to define the parameter 
\begin{equation}\label{eq:alpha}
\alpha\equiv \sqrt{1-\frac{24\,h_H^{[0]}}{c}}=\sqrt{\eta^4-4\tilde{n}(1-\eta^2)}\,,
\end{equation}
which is related to the left temperature of the black hole with the same charges. Note that $\alpha$ becomes imaginary precisely when the heavy state enters the black hole regime \eqref{eq:bhregeta} and that for $\tilde n=1$ and $\eta$ small one has
\begin{equation}\label{eq:alphapert}
\alpha=2 i \left(1-\frac{\eta^2}{2}-\frac{\eta^4}{4}\right)+O(\eta^6)\,,
\end{equation}
where one recognises the function of $\eta$ that appears in the first square bracket in \eqref{eq:conf01}. The large $c$ conformal block with two external heavy states characterised by $\alpha$, two external light states of dimension $h_L=\frac{\Delta}{2}$ and an exchanged primary of dimension $h$ on the cylinder is
\begin{equation}
\mathcal{V}_h(\alpha,\Delta;v)=\left(\frac{\alpha}{2\,\sin\left(\frac{\alpha\,v}{2}\right)}\right)^{\!\Delta}\,\left(i\frac{1-e^{i\alpha v}}{\alpha}\right)^{\!h}\,{}_2F_1(h,h,2h;1-e^{i\alpha v})\,.
\end{equation}
Since our light states have vanishing R-charge, the $SU(2)$ descendants do not contribute and the Virasoro block suffices to capture the whole (bosonic) tower of descendants of an affine primary. An analogous story applies to the right sector, where the parameter $\bar \alpha$ characterising the heavy state can be obtained from \eqref{eq:alpha} by setting $\tilde n=0$:
\begin{equation}
\bar \alpha = \eta^2\,.
\end{equation}
The right-moving block is thus
\begin{equation}
\overline{\mathcal{V}}_{\bar h}(\bar\alpha,\Delta;u)=\left(\frac{\eta^2}{2\,\sin\left(\frac{\eta^2\,u}{2}\right)}\right)^{\!\Delta}\,\left(i\frac{1-e^{i\eta^2 u}}{\eta^2}\right)^{\!h}\,{}_2F_1(h,h,2h;1-e^{i\eta^2 u})\,.
\end{equation}
As we are working perturbatively in $u$ and $\eta$, we just need the leading term of the expansion of the right-moving blocks
\begin{equation}
  \overline{\mathcal{V}}_{\bar h}(\bar\alpha,\Delta;u)\approx u^{-\Delta+\bar h}\,. 
\end{equation}
As for the left-moving blocks, we just need to implement the $\eta$-expansion (which simplifies $\alpha$ as in \eqref{eq:alphapert}), but we are left with a non-trivial $v$-dependence, which will provide a quite stringent check of the supergravity results in \eqref{eq:conf01} and \eqref{eq:conflog}. 

Remarkably, the meromorphic term \eqref{eq:conf01} rewrites as a linear combination of only three blocks:
\begin{equation}\label{eq:block01}
\begin{aligned}
\mathcal{C}(u,v)&=\frac{\mathcal{V}_0(\alpha,\Delta;v)}{u^\Delta}+\frac{\Delta (2-\Delta) \eta^2(1-\eta^2)}{12(\Delta-1)}\frac{\mathcal{V}_1(\alpha,\Delta;v)}{u^{\Delta-1}}+\\ &+\frac{\Delta(\Delta+1)\eta^2(1+\eta^2)}{90(\Delta-1)}\frac{\mathcal{V}_3(\alpha,\Delta;v)}{u^{\Delta-1}}+O(\eta^6)\,,
\end{aligned}
\end{equation}
corresponding to the exchange of primaries of dimension $(0,0)$, $(1,1)$ and $(3,1)$. The logarithmic part of the $\Delta=2$ correlator in \eqref{eq:conflog} contains instead an infinite series of primaries of dimension $(2h,2)$ ($h=1,2,\ldots$)
\begin{equation}\label{eq:blocklog}
\mathcal{C}(u,v)|_{\log}^{\Delta=2}=\eta^4\,\log u\,\sum_{h=1}^\infty c_h \,\mathcal{V}_{2h}(\alpha,2;v)+O(\eta^6)\,,
\end{equation}
with the coefficients of the first few terms being $c_1=\frac{1}{20}$, $c_2=-\frac{32}{525}$, $c_3=-\frac{17}{3780}$, $c_4=-\frac{2}{6435}$.

To interpret these results and verify their consistency with the CFT, it is useful to recall some general features of the HHLL correlators in the large $c$ expansion.

(i) The leading term of the correlator for $c\to \infty$ comes from disconnected Witten diagrams that effectively compute the product of two 2-point functions. This leading contribution can hence be derived from generalized free fields and one can easily estimate that it is proportional to the average number of single-particle constituents of $O_H$, $\bar p$. In the HHLL regime, in which $\bar p\sim N$, the leading term of the correletor thus scales like $N$ and it captures the exchange of multi-particle states with an arbitrary number of traces. While in the LLLL correlators only double-particle states contribute at leading order in $N$, since every extra trace is suppressed by a factor $1/N$, this suppression is compensated by a factor $\bar p$ in the HHLL case, so that the effective expansion parameter is $\frac{\bar p}{N}=1-\eta^2$. Note that we are not claiming that these combinatorial factors give the full $\eta$-dependence of the correlator: depending on the particular form of $O_H$, $O_L$ and of the exchanged operator there might be extra factors of $\eta$ and, of course, certain multi-particle exchanges might be entirely missing. For our choice of $O_H$ and $O_L$, the leading term in $N$ of $\mathcal{C}$ vanishes. This does not mean that the 3-point coupling associated with each exchanged operator in a certain channel vanish at order $N$, but only that all the contributions to the correlator of a given dimension sum up to zero.

(ii) The next order contribution scales like $N^0$ in the HHLL correlator and is captured by the supergravity calculation we have described in Section~\ref{sec:strongcorr}. The Laurent series of this $O(N^0)$ correlator in a certain OPE channel encodes the leading order 3-point couplings of protected single-particle operators and the $1/N$ corrections to the multi-particle states discussed in (i). The single-particles can appear only up to a dimension equal to twice the dimension of $O_L$, after which only multi-particles are exchanged. Moreover there could be logarithmic terms, whose coefficients are proportional to the product of the $1/N$ anomalous dimensions of the multi-particle states with their 3-point couplings at $O(N)$. If there is more than one primary with the same bare dimension the coefficients of the OPE contain a sum over the degenerate states.

(iii) At the order in the large $c$ expansion considered in (ii), the inversion formula of \cite{Caron-Huot:2017vep} links the 3-point couplings and anomalous dimensions of the $p$-particle operators exchanged in one channel with those of the $p-1$-particle operators exchanged in the crossed channels, but only for spin greater than two. Thus, for example, the OPE data of two-particle states with spin greater than two can be entirely derived from the protected data of the single-particles exchanged in the crossed channels.

We can now check if the correlator derived by our supergravity calculation is consistent with the CFT structure reviewed above. We start from the meromorphic piece \eqref{eq:block01}. The term proportional to $\mathcal{V}_0$ represents, of course, the contribution of the Virasoro descendants of the identity and it is the only term that survives in the purely thermal propagator derived from the extremal BTZ geometry. The remaining terms correct the thermal propagator for small but non-vanishing $\eta$. The fact that the correlator at order $O(u^{-\Delta+1})$ contains only states of spin smaller than or equal to two is in agreement with property (iii): At this order, by a straightforward constraint of dimension, the exchanged operators could only be single or double-particle; the supergravity single-particles obviously have spin not greater than two and, by property (iii), the double-particles of spin greater than two are determined by the single-particles exchanged in the crossed channel where $O_H$ fuses with $O_L$ (this is the channel considered in \eqref{eq:OHOLOPE}); but since we have chosen $O_H$ and $O_L$ in different tensor multiplets, their OPE does not contain any single-trace. This argument leaves open the possibilities that the primaries of dimension $(1,1)$ and $(3,1)$ appearing in \eqref{eq:block01} be single or double-particle states. The absence of logarithmic terms at this order in the $u\to 0$ expansion suggests that these primaries are protected single-trace operators, and the $SO(N_f)$ flavour symmetry discussed in Section~\ref{sec:lightstates} requires that the single-traces sit in the gravity multiplet. Thus there is only one possible identification: the $(1,1)$ primary has to be the CPO $\sigma_2$ and the $(3,1)$ primary the $(G^-)^2$ descendant of the CPO $V^+_2$. This however cannot be correct when $\Delta=2$. To understand this point it is useful to remember that the light operator $O_L$ is a descendant of the CPO $s^{(f)}_{\Delta-1}$: $O_L=G^-{\tilde G}^- s^{(f)}_{\Delta-1}$. Hence the correlator $\mathcal{C}$ inherits via the supersymmetric Ward identity some of the properties of the 4-point function involving the CPO's $s^{(f)}_{\Delta-1}$. In particular, as recalled in the point (ii), in both correlators the $u\to 0$ OPE channel can contain single-trace operators only with dimension strictly smaller than twice the dimension of $s^{(f)}_{\Delta-1}$, that is, $2(\Delta-1)$. It follows that for $\Delta=2$ the exchanged operators with $\bar h=1$ have to be double-particles. Since the coefficient of $\mathcal{V}_1$ in \eqref{eq:block01} vanishes for $\Delta=2$, the natural double-trace, $\sigma^{(f')}_1 \bar \sigma^{(f')}_1$, of dimension $(1,1)$ actually does not appear, but there is a contribution of dimension $(3,1)$, that can only be identified with the double-trace $V^+_1 \bar V^+_1$. The latter is a non-supersymmetric state, and its anomalous dimension likely does not vanish, but this does not generate a log in the supergravity correlator because the 3-point coupling between this double-trace and our heavy operators vanishes at $O(N)$. 

We finally analyse the logarithmic term with $\Delta=2$, whose block decomposition is given in \eqref{eq:blocklog}. The operators contributing to this term are non-supersymmetric and thus multi-particle. A first basic, but still non-trivial, consistency check is the absence from \eqref{eq:blocklog} of the vacuum block $\mathcal{V}_0$. A second qualitative feature of \eqref{eq:blocklog} is the presence of an infinite tower of primaries of dimensions $(2h,2)$ for $h=1,\dots,\infty$. Keeping in mind the consequences of the inversion formula recalled in (iii) and the absence of single-particles in the crossed (heavy-light) channels, we should conclude that, at least from $h=3$ onward, these primaries are $p$-particle states with $p>2$. At every level there are actually several degenerate primaries and the exchanged operator is a linear superposition of them. For example at level $(2,2)$ there are the double-particles $O_L \bar O_L$, $s_1^{(f)} \partial\bar\partial \bar s_1^{(f)}$ and the four-particle operators $(s_1^{(f)})^2 (\bar s_1^{(f)})^2$, $s_1^{(f)} \bar s_1^{(f)} J_3 {\tilde J}_3$ and it is not difficult to verify withing the generalized free field theory that a linear combination of these primaries has a 3-point coupling with the two heavy operators $O_H$ and $\bar O_H$ whose $\eta$-dependence goes like $\eta^4$ for $\eta\to 0$, like the one observed in \eqref{eq:blocklog}. Given the non-protected nature of the primaries, it is difficult to perform more quantitative checks, but the qualitative structure of \eqref{eq:blocklog} is consistent with the general CFT expectation 

\section{Summary and Outlook}
\label{sec:concl}
The BTZ propagator, that we recall in \eqref{eq:RBTZmomentum}, cannot represent a two-point correlator in any unitary CFT with a finite-dimensional Hilbert space: its complex poles imply, for example, that the propagator in configuration space decays for large Lorentzian time. Long ago Maldacena \cite{Maldacena:2001kr} pointed out that this is one of the incarnations of the information paradox. In Eq. \eqref{eq:largeLG} we present a deformation of the BTZ propagator that is consistent with unitarity. 
The propagator in Eq.\eqref{eq:largeLG}, represents the two-point correlator of gravity operators
of dimension $\Delta$ in a family of heavy states, $O_H$, representing microstates of the supersymmetric black hole with a non-vanishing left-moving temperature, in the supergravity limit of large $c$ and strong coupling. 

To clarify the meaning and the limitations of our result, it is useful to recall a few facts about the states $O_H$. They are a subfamily of the so called ``graviton gas", coherent superpositions of products of a large number (of order $c$) of copies of supergravity operators. As such, these states are not typical representatives of the statistical ensemble associated with the black hole. On the other hand, this is what allows for an explicit supergravity description of these states in terms of a horizonless geometry with the same charges as the black hole, and ultimately enables us to compute the propagator using standard holographic techniques. One might thus question if our result has any relevance for the black hole problem. For this purpose one should recall that in a particular limit ($\eta\to 0$) the state $O_H$, though still atypical, shares a property that characterizes typical black hole states: the expectation values of all the supergravity operators in this heavy state vanish, and hence at the supergravity level the pure state $O_H$ is indistinguishable from the mixed state represented by the black hole. In this sense we can think of the state at $\eta=0$ as a quantum black hole microstate. Correspondingly the horizonless geometry dual to $O_H$ degenerates for $\eta\to 0$ to the BTZ black hole. 

In conclusion, one can think of the small $\eta$ limit as a way of adding graviton hair to a quantum black hole state. Hence by studying the evolution of the analytic structure of the pure state propagator as $\eta\to 0$ we can understand very explicitly how the non-unitary behaviour of the BTZ propagator arises as a degeneration of a unitary structure. The basic mechanism that emerges from our analysis is, a posteriori, quite natural. As one approaches the black hole limit, the real poles of the unitary propagator become dense and, once the separation between the poles cannot be resolved, they can effectively be replaced by a cut. The discontinuity of the propagator across the cut has complex poles, corresponding to the quasi-normal modes of the black hole. This picture gives a precise meaning to the intuitive expectation that the classical black hole arises from a coarse grained description of its microstates. 

The same physical picture can be stated in the language of algebras \cite{Leutheusser:2021qhd,Leutheusser:2021frk,Leutheusser:2022bgi,Witten:2021jzq,Witten:2021unn,Chandrasekaran:2022eqq}. Defining the large $c$ limit of the algebra of operators of a quantum field theory requires the choice of a heavy state. When the state is our $O_H$ with $\eta\not=0$, the properties of the propagator suggest that the algebra is of type I, even in the strict large $c$ limit, while it degenerates to a type III algebra when $\eta\to 0$. This is completely analogous to what happens for the thermofield double state when one goes from finite $c$ to infinite $c$. 

One of the outcomes of our study is the qualitative similarity of the $\eta\to 0$ limit with the $c\to \infty$ limit. We have just pointed out the analogy at the level of algebras, but the same picture emerges from the analytic structure of the propagator: the appearance of an approximate continuum of poles, giving rise to a cut, and of isolated complex poles after crossing the cut, is what one expects in the large $c$ limit of a typical black hole state, as explained for example in \cite{Dodelson:2022eiz}. The precise reason for this qualitative agreement is unclear to us: maybe, despite our heavy states being atypical, their perturbation by the light operators explores a sizeable fraction of the phase space, or, equivalently, the heavy-light OPE channel of our correlator contains enough states to reproduce, qualitatively, the behaviour expected in a more typical microstate. Whatever the reason for it, one could exploit this analogy to formulate some speculative ideas. The holy grail of theoretical black hole physics is providing an explicitly unitary description of typical black hole states in the strong coupling CFT regime, where the curvature length scales are macroscopic. Typical black hole states lie outside the graviton gas, are described in supergravity by the naive black hole geometry and the corrections to the classical picture that should restore unitarity are expected to be perturbative and non-perturbative $1/c$ effects. In the fuzzball paradigm \cite{Bena:2022rna}, for example, quantum corrections effect the naive black hole geometry at a scale of the order of the horizon radius, even if, generically, the exact quantum state cannot be described by a classical supergravity configuration. This provides a mechanism to restore unitarity of black hole evaporation by introducing order one corrections at the horizon scale~\cite{Mathur:2009hf,Guo:2021blh}. The analogy hinted at above might provide a qualitative guide to describe these quantum corrections, at least at the level of the HHLL correlator, by simply making $\eta^2$ to scale as $e^{-c}$ and assuming that our state $O_H|_{\eta=0}$ is replaced by some typical black hole state. Of course, as we recalled in the Introduction, in this regime our classical approximation breaks down and it is not clear how our results will be modified. 

It might be interesting to further explore the properties of the $\eta\to 0$ limit of our correlator. For example one could reconstruct the spectral form factor from the poles of $G$ and check if it has the broad-brush form -- an initial decay, a ramp and a plateau -- expected for a typical microstate in a unitary theory \cite{Cotler:2016fpe}. And of course it would be especially important to put the analogy between the small $\eta$ and the large $c$ expansions on firmer grounds, for example by verifying if the ansatz for the two-point correlator suggested by this analogy passes some CFT consistency check -- a bootstrap constraint, for instance -- and if such constraints are enough to determine the quantum corrections mentioned above.  In particular the configuration space analysis of Section~\ref{sec:configuration} has revealed the presence of non-trivial primaries exchanged in the LL (or $z_c\to 1$) OPE channel. These primaries, which are both single and multi-trace operators, are crucial to restore the unitarity of the black hole correlator, where, on the contrary, only the identity is exchanged. For our classical state the three-point couplings  involving these non-trivial primaries are controlled by the parameter $\eta$ but for a typical state one expects these couplings to be exponentially suppressed for large $c$. It would be important to understand if the necessity for these extra primaries follows from crossing or some other CFT consistency constraint. Finally, the quantum-corrected two-point correlator in a supersymmetric black hole has been recently computed in \cite{Lin:2022rzw,Lin:2022zxd} by solving the quantum mechanics of the Schwarzian mode of JT gravity. Elucidating the connection between those results and ours will help to improve our understanding of black hole microstates.

\section*{Acknowledgements} 

We would like to thank Giulio Bonelli, Daniel Panea Lichtig and Alessandro Tanzini for collaboration at the initial stages of this project and for feedback during its completion. We would like to thank the participants of the workshops ``String Theory as a bridge between Gauge Theories and Quantum Gravity'' at the University of Turin, ``Recent Developments in Quantum Physics of Black Holes'' at YITP Kyoto University, and ``Black-Hole Microstructure V'' at IPhT Sacly for feedback to a preliminary versions of the results presented in this paper. In particular we would like to thank Camillo Imbimbo, Francisco Morales and Andrei Parnchev for discussions.

SG was supported in part by the MIUR-PRIN contract 2017CC72MK003. RR is partially supported by the UK EPSRC grant ``CFT and Gravity: Heavy States and Black Holes'' EP/W019663/1 and the STFC Consolidated Grants ST/T000686/1. 

\appendix

\section{Connection formulas}\label{app:conn}
We now go through the argument of \cite{Bonelli:2022ten} to compute the reduced confluent Heun connection coefficients $\mathcal{A}$ and $\mathcal{B}$ appearing in the main text. Let us consider the following Liouville correlator (for a review on Liouville theory see \cite{Teschner:2001rv}):
\begin{equation}
\Psi(\Lambda,z)=\langle \Lambda^2 | V_1 (1) \phi_{2,1} (z) | \Delta_0 \rangle \,.
\label{eq:corr}
\end{equation}
Here $V_1(1)$ is a primary operator of weight $\Delta_1$ and $| \Delta_0 \rangle$ is the state created by the operator $V_0(0)$ of weight $\Delta_0$. We parametrize scaling dimensions in terms of Liouville momenta $\alpha_i$
\begin{equation}
\Delta_i = \frac{Q^2}{4}-\alpha_i^2 \,, \quad Q = b+\frac{1}{b} \,,
\end{equation}
where $Q$ parametrizes the central charge $c = 1+6Q^2$. The state $\langle \Lambda |$ is obtained by inserting at infinity a so called rank 1/2 irregular state. It is characterized by 
\begin{equation}
\begin{aligned}
&\langle \Lambda^2 | L_0 = \Lambda^2 \partial_{\Lambda^2} \langle \Lambda^2 | \,, \\ &\langle \Lambda^2 | L_{-n} = -\frac{\Lambda^2}{4} \delta_{1, n} \langle \Lambda^2 | \,, n\ge 1 \,.
\end{aligned}
\end{equation}
Note that the action of $L_0$ fixes
\begin{equation}
\langle \Lambda^2 | \Delta \rangle = C_\alpha | \Lambda^2 |^{2\Delta} \,,
\label{eq:overlap}
\end{equation}
Such a state can be formally obtained as a confluence limit of two primary operators \cite{Gaiotto:2009ma,Gaiotto:2012sf,Marshakov:2009gn}. Accordingly, $C_\alpha$ in \eqref{eq:overlap} can be obtained as a collision limit of a three point function. Finally, $\phi_{2,1}$ is the degenerate state of weight $\Delta_{2,1} = -\frac{1}{2}-\frac{3}{4} b^2$. It satisfies
\begin{equation}
\left(b^{-2} L_{-1}^2 (z) + L_{-2} (z) \right) \phi_{2,1} (z) = 0 \,,
\label{eq:nulldesc}
\end{equation}
meaning that the descendant obtained by this combination of Virasoro generators has zero norm. Inserting \eqref{eq:nulldesc} inside \eqref{eq:corr} we find 
\begin{equation}
\left(b^{-2} \partial_z^2 - \frac{2z-1}{z(z-1)} \partial_z + \frac{\Lambda^2 \partial_{\Lambda^2} - \Delta_{2,1}-\Delta_1-\Delta_0}{z(z-1)} + \frac{\Delta_1}{(z-1)^2} + \frac{\Delta_0}{z^2} - \frac{\Lambda^2}{4z}\right) \Psi(\Lambda,z) = 0\,.
\label{eq:bpz}
\end{equation}
Now the argument will proceed as follows: we will first show how equation \eqref{eq:bpz} reduces to a reduced confluent Heun equation of the form \eqref{eq:RCHE} when we take the semiclassical limit $c\to \infty$, and then we will compute the connection coefficients exploiting crossing symmetry of \eqref{eq:corr}.

\subsection{Regular and irregular conformal blocks}
Equation \eqref{eq:bpz} being second order in $z$ reflects the fact that in the OPE between $\phi_{2,1}$ and a primary only two operators are exchanged. One has
\begin{equation}\label{eq:a7}
\begin{aligned}
&\phi_{2,1}(z) V_i (z_i) \sim \sum_{\pm, n} (z-z_i)^{\frac{bQ}{2}\pm b \alpha_i+n} V^{(n)}_{i\pm}(z_i) \,, \\
&\Delta_{i\pm} = \frac{Q^2}{4}-\left(\alpha_i \mp \frac{b}{2}\right)^2 \,,
\end{aligned}
\end{equation}
where $V^{(n)}$ is the $n$-th Virasoro descendant of $V$. Here for simplicity we are just restricting to the holomorphic part of the OPE. To be more precise one should consider the antiholomorphic part as well and weight the sum with the three point functions.

OPE with the irregular state is more complicated. One can expand the result either in Verma modules or in terms of irregular states. For the first case one can insert an identity in a given Verma module 
\begin{equation}
\begin{aligned}
\langle \Lambda^2 | V_i(t) &\sim \sum_n \langle \Lambda^2 | \prod_{Y,Y' | |Y|=|Y'|=n} L_{-Y} | \Delta \rangle \left( \langle \Delta | L_Y L_{-Y'} | \Delta \rangle \right)^{-1} \langle \Delta | L_Y' V_i(t) = \\ &= \langle \Delta | \left( 1 - \frac{1}{2\Delta} \frac{\Lambda^2}{4} L_1 + \mathcal{O}\left(\Lambda^4\right) \right) V_i(t) \,,
\end{aligned}
\label{eq:regope}
\end{equation}
where $Y,Y'$ are Young tableaux of length respectively $|Y|,|Y'|$. Otherwise one can write the following ansatz
\begin{equation}\label{eq:irregope}
\langle \Lambda^2| V_i(t) = \sum_{k=0}^\infty (\Lambda^2)^\lambda t^{\tau-\frac{k}{2}} e^{\gamma \Lambda \sqrt{t}} \langle \Lambda^2;\frac{k}{2}| \,,
\end{equation}
and constrain the coefficients imposing covariance under Virasoro generators. Here $\langle \Lambda^2;\frac{k}{2}|$ are the irregular descendant of the irregular state. In our case, the two approaches are useful to study different regimes: the first one applies to the limit $\eta\to 1$, while the second one is relevant to the limit $\eta\to 0$. In principle we should include an integral in both~\eqref{eq:regope} (over $\Delta$ or $\alpha$ again defined as in~\eqref{eq:a7}) and in~\eqref{eq:irregope} (over $\gamma$), but we do not need to write them explicitly as we will work at the level of the integrands. In practice we use the following perturbative expansion of~\eqref{eq:irregope}
\begin{equation}
\begin{aligned}
&\langle \Lambda^2 ; \frac{1}{2}| = \left( c_1 \Lambda^{-1} + c_2 \partial_\Lambda \right) \langle\Lambda^2| \,, \\
&\langle \Lambda^2 ; 1| = \left( s_1 \Lambda^{-2} + s_2 \partial_{\Lambda^2} + s_4 \partial_\Lambda^2 \right) \langle\Lambda^2| + s_3 \langle\Lambda^2| L_{1} \,, \\
&\langle \Lambda^2 ; \frac{3}{2} | = \dots 
\end{aligned}
\end{equation}
where one finds \cite{Nagoya:2018pgp}
\begin{equation}
\begin{aligned}
&\lambda = \frac{\gamma^2}{8}+\frac{\Delta_i}{2} \,, \\
&\tau = \frac{\gamma^2}{8}-\frac{\Delta_i}{2} \,, \\
&c_1 = \frac{\gamma}{16} \left(\gamma^2 + 4 \Delta_i - 2 - 6 Q^2\right) \,, \\
&c_2 = \gamma \,, \\
&s_1 = \frac{\gamma^6}{512}-\frac{\gamma^4(c-4\Delta_i+6)}{256} + \\ &\quad +
\frac{\gamma^2}{1536} \left( 3c^2-24 \Delta_i c+74c + 48 \Delta_i^2 - 168 \Delta_i + 103 \right) - \frac{\Delta_i(\Delta_i-c-2)}{16} \,, \\
&s_2 = \frac{1}{8}\left(\gamma^4-16\Delta_i+\gamma^2\left(-6-6Q^2+4 \Delta_i\right)\right) \,, \, s_3=0 \,, \\
&s_4 = \frac{\gamma^2}{2} \,.
\end{aligned}
\label{eq:irregope2}
\end{equation}
When $V_i = \phi_{2,1}$ we get $\gamma = \gamma_\pm = \pm b$. We will refer to \eqref{eq:irregope2} as \textit{irregular OPE}. Note that while \eqref{eq:regope} will result in a series in $\Lambda^{2n}$ when inserted in correlators, \eqref{eq:irregope2} will result in a series in powers of $\Lambda^{-1}$. 

Let us now consider the conformal blocks expansion of \eqref{eq:corr}. Schematically
\begin{equation}
  \langle \Lambda^2 | V_1 (1) \phi_{2,1} (z) | \Delta_0 \rangle = 
  \left( \text{3 pt functions} \right) |f
  \left(\Lambda, z\right)| \,,
\end{equation}
where again we do not explicitly write the integral over $\alpha$ or $\gamma$. 
Here $f$ is the so called conformal block: we will call $f$ a regular conformal block every time we consider the OPE \eqref{eq:regope} with regular states and an irregular conformal block otherwise.  

In the semiclassical limit
\begin{equation}
\alpha_i = a_i / b \,, \quad \Lambda = L/b \,, \quad \gamma = g / b \,, ~ \mbox{ or } ~ \alpha = a/b\,, \quad b \to 0 \,.
\end{equation}
the conformal blocks behave as follows \cite{Zamolodchikov:1995aa}
\begin{equation}\label{eq:a14}
f(\Lambda, z) = e^{\frac{1}{b^2} \tilde{F}(L) + W\left(L, z\right)} \,.
\end{equation}
Here all the $z$ dependence appears at a subleading order in $b$ since $\Delta_{2,1} / \Delta_i \to 0$ as $b \to 0$. In particular to compute $\tilde{F}$ we only need the correlator
\begin{equation}
\langle \Lambda^2 | V_1 (1)| \Delta_0 \rangle = e^{\frac{1}{b^2} \tilde{F}\left(L\right)} \,.
\label{eq:Ftilde}
\end{equation}
According to the AGT duality when no irregular OPEs are involved the correlator \eqref{eq:Ftilde} is equal to the partition function of a weak coupled $\mathcal{N}=2$ $SU(2)$ gauge theory with 2 hypermultiplets in the fundamental representation of the gauge group. This partition function can be computed by localization, and the result is expressed as a combinatorial series. In the following we will denote $\tilde{F} = \left(\frac{1}{4}-a^2\right) \log L^2 + F$, where $a$ is the semiclassical Liouville momentum of the intermediate operator, when no irregular OPE is involved, $\tilde{F} \sim F_D$ otherwise\footnote{$F_D$ stands for $F$ dual. In fact the $1/L$ expansion of $\tilde{F}$ corresponds to an expansion at strong coupling in the gauge theory, related to the weak coupling expansion by S duality.}. The combinatorial expression for $F$ reads \cite{Flume:2002az,Bruzzo:2002xf}
\begin{equation}\label{eq:Fcomb}
F\left(L^2\right) = \lim_{b\to 0} b^2 \log \sum_{\vec{Y}} \left(\frac{\Lambda^2}{4}\right)^{|\vec{Y}|} z_{vec} \left( \vec{\alpha},\vec{Y} \right) \prod_{\theta = \pm} z_{hyp} \left(\vec{\alpha},\vec{Y}, \alpha_1+\theta \alpha_0\right) \,,
\end{equation}
where $\vec{\alpha} = \left(\alpha, - \alpha\right)$, $\vec{Y} = \left(Y_1,Y_2\right)$ is a pair of Young tableaux of length $|Y_1|, |Y_2|$, and
\begin{equation}
\begin{aligned}
&z_{hyp} \left(\vec{\alpha}, \vec{Y}, \mu\right) = \prod_{k=1,2} \prod_{(i,j) \in Y_k} \left(\alpha_k+\mu+b^{-1}\left(i-\frac{1}{2}\right)+b\left(j-\frac{1}{2}\right)\right) \,, \\
&z_{vec} = \prod_{k = 1,2} \prod_{(i,j) \in Y_k} E^{-1} \left(\alpha_k-\alpha_l,Y_k,Y_l,(i,j)\right) \prod_{(i',j') \in Y_l} \left(Q - E \left(\alpha_l-\alpha_k,Y_l,Y_k,(i',j')\right)\right)^{-1} \,, \\
&E\left(\alpha,Y_1,Y_2,(i,j)\right) = \alpha-b^{-1} L_{Y_2}((i,j)) + b\left(A_{Y_1}((i,j))+1\right) \,.
\end{aligned}
\end{equation}
$L_Y ((i, j)), A_Y ((i, j))$ denote respectively the leg-length and the arm-length of the box at the site $(i, j)$ of the tableau $Y$, meaning that given $Y = (\nu_1' \ge \nu_2' \ge \dots)$ and its transpose $Y^T = (\nu_1 \ge \nu_2 \ge \dots)$, 
\begin{equation}
A_Y((i,j)) = \nu_i'-j \,, \quad L_Y((i,j)) = \nu_j-i \,.
\end{equation}
For concreteness, at leading order one finds 
\begin{equation}
F = \frac{-1+4a^2-4a_0^2+4a_1^2}{8-32a^2} L^2 + \mathcal{O}(L^4) \,.
\end{equation}
On the other hand, $F_D$ does not have such a simple combinatorial expression. Nevertheless, it can be computed order by order in $L^{-1}$ performing the irregular OPE. We have at subleading order
\begin{align}
&\langle \Lambda^2 | V_1(1) | \Delta_0 \rangle = \sum_{k=0}^\infty \left(\Lambda^2\right)^\lambda t^{\tau-\frac{k}{2}} e^{\gamma \Lambda \sqrt{t}} \langle \Lambda^2 ; \frac{k}{2}| \Delta_0 \rangle\Big|_{t=1} = \\ \nonumber  & = \Lambda^{\frac{\gamma^2}{4}+\Delta_1+2\Delta_0} e^{\gamma \Lambda} \left[ 1+\frac{\gamma}{16b^2\Lambda}\left(2+2b^4+b^2\left(2-32\alpha_0^2+\gamma^2+4\Delta_1\right)\right) + \mathcal{O}(\Lambda^{-2})\right] \,.
\end{align}
Again from \eqref{eq:Ftilde} (where we now stop at sub-subleading order for further convenience)
\begin{equation}
\begin{aligned}
F_D &\simeq \frac{g\left(3-32a_0^2-4a_1^2+g^2\right)}{16L} + \\ &+\frac{-9-16a_1^4-34g^2-5g^4+8a_1^2(5+3g^2)+128a_0^2(1-4a_1^2+3g^2)}{256L^2} \,.
\label{eq:FD}
\end{aligned}
\end{equation}
Note that while $F$ is a series in $L^2$, $F_D$ is a series in $L^{-1}$.
Coming back to \eqref{eq:bpz}, the derivative with respect to $\Lambda$ decouples at leading order in $b$, since in this limit we have
\begin{equation}
  \Lambda^2 \partial_{\Lambda^2} f
  \left(\Lambda, z\right) \simeq f
  \left(\Lambda, z\right) b^{-2} L^2 \partial_{L^2} \tilde{F}\left(L\right) = b^{-2} u f
  \left(\Lambda, z\right) \,,
\end{equation}
where we used~\eqref{eq:a14} and the Matone relation $u = L^2 \partial_{L^2} \tilde{F}$. In the mathematical literature $u$ is the so called accessory parameter. Since $u$ is completely independent of $z$, in this limit \eqref{eq:bpz} turns into an ODE:
\begin{equation}
\left(\partial_z^2 + \frac{u + a_0^2 + a_1^2 - \frac{1}{2}}{z(z-1)} + \frac{\frac{1}{4}-a_1^2}{(z-1)^2} + \frac{\frac{1}{4}-a_0^2}{z^2} - \frac{L^2}{4z}\right) e^{W(L, z)} = 0\,.
\end{equation}
This equation has regular singularities at $z=0,1$, and an irregular singularity of rank $1/2$ at infinity: it is the reduced confluent Heun equation.

\subsection{Three point functions and connection coefficients}
We now derive the connection formulas of interest. Let us start with the case involving only regular conformal blocks. The first step is to write down the crossing symmetry constraints relating conformal blocks expansion for $z \sim 0$ and $z \sim 1$. In both cases we keep $L^2 \ll 1$, that is we always consider regular OPE with the irregular state. We denote schematically the resulting conformal blocks as 
\begin{equation}
\mathfrak{F}_{\alpha_{0\theta}}^\alpha \left(\Lambda^2, z\right) \,, \quad \mathfrak{F}_{\alpha_{1\theta}}^\alpha \left(\Lambda^2, 1-z\right) \,.
\end{equation}
The constraint reads
\begin{equation}
\begin{aligned}
\langle \Lambda^2 | V_1(1) \phi_{2,1}(z) | \Delta_0 \rangle &= \int d \alpha \, C_\alpha \sum_{\theta = \pm} C_{\alpha_{2,1} \alpha_0}^{\alpha_{0 \theta}} C_{\alpha_1 \alpha_{0 \theta}}^\alpha \bigg| \mathfrak{F}_{\alpha_{0\theta}}^\alpha \left(\Lambda^2, z\right) \bigg|^2 = \\ &= \int d \alpha \, C_\alpha \sum_{\theta = \pm} C_{\alpha_{2,1} \alpha_1}^{\alpha_{1 \theta}} C_{\alpha_0 \alpha_{1 \theta}}^\alpha \bigg| \mathfrak{F}_{\alpha_{1\theta}}^\alpha \left(\Lambda^2, 1-z\right) \bigg|^2\,.
\end{aligned}
\label{eq:regcrossing}
\end{equation}
Here the $C$'s are the OPE coefficients and three point functions. They are given by
\begin{equation}
\begin{aligned}
&C_\alpha = 2^{-4\Delta} e^{-2\pi i \Delta} \Upsilon_b \left(Q+2\alpha\right) \,, \\ 
&C_{\alpha_{2,1} \alpha}^{\alpha_+} = 1 \,, \\
&C_{\alpha_{2,1} \alpha}^{\alpha_-} = b^{2bQ} \frac{\gamma(2 b \alpha)}{\gamma\left(bQ+2b\alpha\right)} \,, \\
&C_{\alpha_1 \alpha_2 \alpha_3} = \frac{\Upsilon_b'(0) \Upsilon_b\left(Q+2\alpha_1\right)\Upsilon_b\left(Q+2\alpha_2\right)\Upsilon_b\left(Q+2\alpha_3\right)}{\Upsilon_b\left(\frac{Q}{2}+\alpha_1+\alpha_2+\alpha_3\right)\Upsilon_b\left(\frac{Q}{2}+\alpha_1+\alpha_2-\alpha_3\right)} \times \\ &\quad\quad \times \frac{1}{\Upsilon_b\left(\frac{Q}{2}+\alpha_1-\alpha_2+\alpha_3\right)\Upsilon_b\left(\frac{Q}{2}-\alpha_1+\alpha_2+\alpha_3\right)} \,, \\
&G_\alpha = \frac{\Upsilon_b\left(Q+2\alpha\right)}{\Upsilon_b\left(2\alpha\right)} \,, \\
&C_{\alpha_2 \alpha_3}^{\alpha_1} = G_{\alpha_1}^{-1} C_{\alpha_1 \alpha_2 \alpha_3} \,.
\end{aligned}
\end{equation}
We will not define carefully the special function $\Upsilon_b$, we instead refer to \cite{Teschner:2001rv}. We only report its most important property, namely
\begin{equation}
\Upsilon_b (x+b) = \gamma(bx) b^{1-2bx} \Upsilon_b(x) \,, \quad \gamma(x) = \frac{\Gamma(x)}{\Gamma(1-x)} \,.
\end{equation}
$C_{\alpha_1 \alpha_2 \alpha_3} = \langle V_1(\infty) V_2(1) V_3(0) \rangle$ is the DOZZ three point function and $C_{\alpha_2 \alpha_3}^{\alpha_1}$ the corresponding OPE coefficient. Assuming that \eqref{eq:regcrossing} holds term by term in the integral we get
\begin{equation}
\sum_{\theta = \pm} C_{\alpha_{2,1} \alpha_0}^{\alpha_{0 \theta}} C_{\alpha_1 \alpha_{0 \theta}}^\alpha \bigg| \mathfrak{F}_{\alpha_{0\theta}}^\alpha \left(\Lambda^2, z\right) \bigg|^2 = \sum_{\theta' = \pm} C_{\alpha_{2,1} \alpha_1}^{\alpha_{1 \theta'}} C_{\alpha_0 \alpha_{1 \theta'}}^\alpha \bigg| \mathfrak{F}_{\alpha_{1\theta'}}^\alpha \left(\Lambda^2, 1-z\right) \bigg|^2 \,.
\end{equation}
We can now write the ansatz
\begin{equation}
\mathfrak{F}_{\alpha_{0\theta}}^\alpha \left(\Lambda^2, z\right) = \sum_{\theta \theta'} M_{\theta \theta'} \mathfrak{F}_{\alpha_{1\theta'}}^\alpha \left(\Lambda^2, 1-z\right)
\label{eq:hypcross}
\end{equation}
and try to solve for the connection matrix $M$. This might look complicated, but it is not. In fact, in \eqref{eq:hypcross} only regular OPE coefficients enter: this equation knows nothing about the irregular state. In particular, this is the same crossing symmetry constraint that one would get from 
\begin{equation}
\langle \Delta | V_1 (1) \phi_{2,1}(z) | \Delta_0 \rangle \,.
\end{equation}
Since the conformal blocks of this correlator are just hypergeometric functions, whose connection coefficients are well known, we can read off $M$. We find
\begin{equation}
M_{\theta \theta'} = \frac{\Gamma\left(-2\theta' b\alpha_1\right)\Gamma\left(1+2\theta b \alpha_0\right)}{\Gamma\left(\frac{1}{2}-\theta' b \alpha_1 + \theta b \alpha_0 + b \alpha_\infty\right)\Gamma\left(\frac{1}{2}-\theta' b \alpha_1 + \theta b \alpha_0 - b \alpha_\infty\right)} \,.
\end{equation}
The last step is to take the semiclassical limit of equation \eqref{eq:hypcross}. Let us call $\mathcal{F}$ the semiclassical conformal blocks. We find
\begin{equation}
\mathcal{F}_{\alpha_0\theta}^\alpha = \lim_{b\to0} e^{-\frac{1}{b^2}F}\mathfrak{F}_{\alpha_0\theta}^\alpha = e^{-\frac{\theta}{2}\partial_{a_0} F(L^2)} z^{\frac{1}{2}+\theta a_0} \left(1+\mathcal{O}(z)\right) \,.
\end{equation}
Finally, this gives for the small $L$ $\mathcal{A}$ and $\mathcal{B}$ in \eqref{eq:boundarybehavior} 
\begin{equation}
\mathcal{A} = M_{+-} e^{\frac{1}{2} \partial_{a_1} F} \,, \quad \mathcal{B} = M_{++} e^{-\frac{1}{2} \partial_{a_1} F} \,.
\end{equation}
As noted in the main text, these connection coefficients will depend on the semiclassical intermediate dimension $a$. On the other hand, in the equation only the accessory parameter $u$ and the external dimensions will appear. However we have seen that
\begin{equation}
u = \frac{1}{4}-a^2 + L^2 \partial_{L^2} F = \frac{1}{4}-a^2 + \frac{-1+4a^2-4a_0^2+4a_1^2}{8-32a^2} L^2 + \mathcal{O}(L^4) \,.
\end{equation}
This is the so called Matone relation \cite{Matone:1995rx,Flume:2004rp}. In order to compute the connection coefficients in terms of parameters entering the equation we need to invert the Matone relation to get $a(u)$. This can be done order by order in $L^2$ by just setting $a = \sum_{n\ge0} A_n L^{2n}$ and solving for the $A_n$'s. For concreteness\footnote{Note that the Matone relation is a second order algebraic equation, therefore it will have two solutions $\pm |a|$. However all the relevant formulas are symmetric in $a \to -a$ so we can choose one of the two solutions without loss of generality.}
\begin{equation}\label{eq:Matonea}
a = - \sqrt{1-4u} + \frac{a_0^2-a_1^2+u}{8u\sqrt{1-4u}}L^2 + \mathcal{O}(L^4) \,.
\end{equation}

We now move to compute the connection coefficient in the large $L$ regime. In this case, if one writes down the crossing symmetry constraint from $z \sim 0$ to $z \sim 1$ as in the previous case, the solution for the connection matrix will not be easy to read off. To find again a solution in terms of known connection coefficients a convenient trick is to proceed in two steps and first solve for the connection matrix from $0$ to $\infty$, and then from $\infty$ to $1$. We will schematically denote the conformal blocks for $z \sim 0, 1, \infty$ respectively as
\begin{equation}
\mathfrak{E}^{(\theta,\gamma)}\left(\Lambda^{-1},z\right)\,,\quad \mathfrak{E}^{(\theta,\gamma)}\left(\Lambda^{-1},1-z\right)\,,\quad \mathfrak{E}^{(\theta,\gamma)}\left(\Lambda^{-1},\frac{1}{\sqrt{z}}\right) \,,
\end{equation}
Let us start with the first step. Crossing symmetry from $0$ to $\infty$ reads
\begin{equation}
\begin{aligned}
\langle \Lambda^2 | V_1 (1) \phi_{2,1} (z) | \Delta_0 \rangle &= \int d\gamma B_{\alpha_1 \gamma} \sum_\theta C_{\alpha_{0\theta}} C^{\alpha_{0\theta}}_{\alpha_{2,1} \alpha_0} \left| \mathfrak{E}^{(\theta, \gamma)}\left(\Lambda^{-1},z\right)\right|^2 = \\ &= \int d\gamma' B_{\alpha_1 \gamma'} \sum_{\theta'} C_{\alpha_{0}} B_{\alpha_{2,1}} \left| \mathfrak{E}^{(\theta',\gamma')}\left(\Lambda^{-1},\frac{1}{\sqrt{z}}\right)\right|^2\,,
\end{aligned}
\label{eq:irrcross0inf}
\end{equation}
where $B_{\alpha \gamma}$ is the OPE coefficient with the irregular state and, and $B_{\alpha_{2,1}}$ the OPE coefficient with the irregular state and $\phi_{2,1}$. They are given by (see equation (A.30) in \cite{Bonelli:2016qwg})
\begin{equation}
\begin{aligned}
&B_{\alpha_{2,1}} = 2^{b^2} e^{i \pi \frac{bQ}{2}} \,, \\
&B_{\gamma \alpha} = 2^{\gamma^2} e^{-i \pi \left(\frac{\gamma^2}{4}+\Delta\right) } \frac{\Upsilon_b\left(Q+2\alpha\right)}{\Upsilon_b \left(\frac{Q+\gamma}{2} + \alpha\right)\Upsilon_b \left(\frac{Q+\gamma}{2} - \alpha\right)} \,.
\end{aligned}
\end{equation}
Again we can assume that \eqref{eq:irrcross0inf} holds term by term in the integral and write
\begin{equation}
\sum_\theta C_{\alpha_{0\theta}} C^{\alpha_{0\theta}}_{\alpha_{2,1} \alpha_0} \left| \mathfrak{E}^{(\theta, \gamma)}\left(\Lambda^{-1},z\right)\right|^2 = \sum_{\theta'} C_{\alpha_{0}} B_{\alpha_{2,1}} \left| \mathfrak{E}^{(\theta',\gamma)}\left(\Lambda^{-1},\frac{1}{\sqrt{z}}\right)\right|^2 \,.
\end{equation}
Then we write
\begin{equation}
\mathfrak{E}^{(\theta, \gamma)}\left(\Lambda^{-1},z\right) = \sum_{\theta'} Q_{\theta \theta'} \mathfrak{E}^{(\theta',\gamma)}\left(\Lambda^{-1},\frac{1}{\sqrt{z}}\right) \,.
\label{eq:step1}
\end{equation}
and solve for the connection matrix $Q_{\theta \theta'}$. Similarly to what happened before, now it's the dependence of $\alpha_1$ drops that out, and we get the same constraint we would have gotten from
\begin{equation}
\langle \Lambda^2 | \phi_{2,1} (z) | \Delta_0 \rangle \,.
\end{equation}
Since this operator satisfies the Bessel equation, again we can read off the solution for $Q$ from the connection coefficients of the Bessel equation. We find
\begin{equation}
Q_{\theta \theta'}(b\alpha_0) = \frac{2^{2\theta b \alpha_0}}{\sqrt{2\pi}} \Gamma\left(1+2\theta b \alpha_0\right) e^{i \pi \frac{1-\theta'}{2}\left(\frac{1}{2}+2\theta b \alpha_0\right)} \,.
\end{equation}
We now move to the second step. Crossing from $1$ to $\infty$ reads
\begin{equation}
\begin{aligned}
\int d\gamma C_{\alpha_0} \sum_\theta B_{\alpha_{1\theta} \gamma} &C^{\alpha_{1\theta}}_{\alpha_{2,1} \alpha_1} \left| \mathfrak{E}^{(\theta,\gamma)}\left(\Lambda^{-1},1-z\right)\right|^2 = \\ = &\int d\gamma C_{\alpha_{0}} B_{\alpha_1 \gamma} B_{\alpha_{2,1}} \sum_\theta \left| \mathfrak{E}^{(\theta,\gamma)}\left(\Lambda^{-1}\frac{1}{\sqrt{z}}\right)\right|^2 \,
\label{eq:strangeconstraint}
\end{aligned}
\end{equation}
Note that, in the notation of \cite{Bonelli:2022ten},
\begin{equation}
B_{\alpha \gamma} = 2^{\gamma^2} e^{- i \pi \frac{\gamma^2}{4}} C_{\frac{\gamma}{2} \alpha}\,, \,\,\,\, B_{\alpha_{2,1}} = 2^{b^2} e^{i \pi \frac{b^2}{4}} e^{- i \pi \theta b \frac{\gamma}{2}} B_{\frac{\gamma}{2} \alpha_{2,1}}^{\left(\frac{\gamma}{2}\right)_\theta} \,,
\end{equation}
where $C_{\gamma/2 \alpha}$ and $B_{\frac{\gamma}{2} \alpha_{2,1}}^{\left(\frac{\gamma}{2}\right)_\theta}$ are the structure constants appearing in fusions of a (degenerate) primary field with a rank 1 irregular state. The idea is to rewrite equation \eqref{eq:strangeconstraint} so that it resembles equation (3.2.26) in \cite{Bonelli:2022ten}, and read off the solution. In the right hand side of \eqref{eq:strangeconstraint} the OPE coefficients however the structure constant do not depend on $\theta$, while in equation (3.2.26) in \cite{Bonelli:2022ten} they do. In order to overcome this problem we can shift the $\gamma$ contour and write
\begin{equation}\label{eq:newcont}
\begin{aligned}
\int d\gamma C_{\alpha_0} \sum_\theta B_{\alpha_{1\theta} \gamma} &C^{\alpha_{1\theta}}_{\alpha_{2,1} \alpha_1} \left| \mathfrak{E}^{(\theta,\gamma)}\left(\Lambda^{-1},1-z\right)\right|^2 = \\ = &\int d\gamma C_{\alpha_{0}} \sum_\theta B_{\alpha_1 \gamma_{-\theta}} B_{\alpha_{2,1}} \left| \mathfrak{E}^{(\theta,\gamma_{-\theta})}\left(\frac{1}{\sqrt{z}}\right)\right|^2 \,,
\end{aligned}
\end{equation}
where $\gamma_\theta = \gamma \mp \theta b$. Now the constraint reads
\begin{equation}
\sum_\theta B_{\alpha_{1\theta} \gamma} C^{\alpha_{1\theta}}_{\alpha_{2,1} \alpha_1} \left| \mathfrak{E}^{(\theta,\gamma)}\left(\Lambda^{-1},1-z\right)\right|^2 = \sum_{\theta'}B_{\alpha_1 \gamma_{-\theta'}} B_{\alpha_{2,1}} \left| \mathfrak{E}^{(\theta', \gamma_{-\theta'})}\left(\frac{1}{\sqrt{z}}\right)\right|^2 \,.
\end{equation}
This can be solved in terms of connection coefficients of confluent hypergeometrics. In particular one finds (neglecting factors of $b^{\#}$ that can be restored in the end simply imposing finiteness of the $b\to0$ limit)
\begin{equation}
2^{- \theta b \gamma + b^2} \mathfrak{E}^{(-\theta, \gamma_\theta)}\left(\Lambda^{-1}, \frac{1}{\sqrt{z}}\right) \sim \sum_{\theta'} \mathcal{N}_{\theta \theta'}^{-1} \left(\frac{b \gamma}{2}, b \alpha_1 \right) \mathfrak{E}^{(\theta', \gamma)} \left(\Lambda^{-1}, 1-z\right) \,,
\label{eq:step2}
\end{equation}
where
\begin{equation}
\mathcal{N}_{\theta \theta'}^{-1} (m,a) = \frac{\Gamma\left(-2\theta' a\right)}{\Gamma\left(\frac{1}{2}+\theta m - \theta' a\right)} e^{i \pi \left(\frac{1+\theta}{2}\right)\left(-\frac{1}{2}-m-\theta' a\right)} \,
\end{equation}
are the connection coefficients of the confluent hypergeometrics. Now we just need to write down the full connection formula concatenating \eqref{eq:step1} and \eqref{eq:step2}. We find
\begin{equation}
\mathfrak{E}^{(\theta,\gamma)} (\Lambda^{-1},z) \sim \sum_{\sigma,\theta'=\pm} 2^{-\sigma b \gamma_{-\sigma}-b^2} \mathcal{Q}_{\theta\sigma}(b\alpha_0)\mathcal{N}^{-1}_{-\sigma \theta'} \left(\frac{b\gamma_{\sigma}}{2},b\alpha_1\right) \mathfrak{E}^{(\theta',\gamma_{\sigma})}(\Lambda^{-1},1-z) \,.
\end{equation}
Finally we take the semiclassical limit of the previous equation. Calling $\mathcal{E}$ the semiclassical blocks we have (restoring the $b^{\#}$ factors)
\begin{equation}
\begin{aligned}
&\mathcal{E}^{(\theta,g)}(L^{-1},z) = \lim_{b\to0} e^{-\frac{1}{b^2}F_D} b^{2\theta b \alpha_0} \mathfrak{E}^{(\theta,\gamma)} = e^{-\frac{\theta}{2}\partial_{a_0}F_D} L^{2\theta a_0}z^{\frac{1}{2}+\theta a_0}\left(1+\mathcal{O}(z)\right) \,, \\
&\mathcal{E}^{(\theta,g_{\sigma})}(L^{-1},1-z) = \lim_{b\to0} e^{-\frac{1}{b^2}F_D} b^{\frac{-\sigma b\gamma_{-\sigma}}{2}+\theta' b \alpha_1} 2^{-\sigma b \gamma_{-\sigma}-b^2} \mathfrak{E}^{(\theta',\gamma_{\sigma})} = \\ &= e^{-\frac{\theta}{2}\partial_{a_1}F_D-\sigma L -\sigma \partial_g F_D} 2^{-\sigma g} L^{\theta a_1 - \frac{\sigma g}{2}} (1-z)^{\frac{1}{2}+\theta' a_1} (1+\mathcal{O}(1-z)) = \\ &= e^{-\sigma L -\sigma \partial_g F_D} (4L)^{-\sigma \frac{g}{2}} \mathcal{E}^{(\theta,g)}(L^{-1},1-z) \,,
\end{aligned}
\end{equation}
where the factor $e^{-\sigma \partial_g F_D}$ comes from the semiclassical limit as in
\begin{equation}
\lim_{b\to0} e^{\frac{F_D(g_\sigma)-F_D(g)}{b^2}} = \lim_{b\to0} e^{\frac{F_D(g-\sigma b^2)-F_D(g)}{b^2}} = e^{-\sigma \partial_g F_D} \,.
\end{equation}
The semiclassical connection formula reads 
\begin{equation}
\mathcal{E}^{(\theta,g)}(L^{-1},z) = \sum_{\sigma,\theta'} \mathcal{Q}_{\theta\sigma}(a_0)\mathcal{N}^{-1}_{-\sigma \theta'} \left(\frac{g}{2},a_1\right) e^{-\sigma L -\sigma \partial_g F_D} (4L)^{-\sigma \frac{g}{2}} \mathcal{E}^{(\theta,g)}(L^{-1},1-z) \,.
\end{equation}
We can finally spell out the connection formula of our interest. Imposing regular boundary condition at $z=0$ selects the $\theta=+1$ solution, and
\begin{equation}\label{eq:almostourconn}
\begin{aligned}
\mathcal{E}^{(+,g)}(z)&=\sum_{\theta'} \bigg( \frac{2^{2a_0}}{\sqrt{2\pi}} \frac{\Gamma\left(1+2a_0\right)\Gamma\left(-2\theta' a_1\right)}{\Gamma\left(\frac{1- g}{2}-\theta' a_1\right)} e^{- L - \partial_g F_D} (4L)^{-\frac{g}{2}} + \\  +& \frac{2^{2a_0}e^{i\pi\left(2a_0-\frac{g}{2} -\theta' a_1\right)}}{\sqrt{2\pi}} \frac{\Gamma\left(1+2a_0\right)\Gamma\left(-2\theta' a_1\right)}{\Gamma\left(\frac{1+ g}{2}-\theta' a_1\right)} e^{ L+\partial_g F_D} (4L)^{\frac{g}{2}} \bigg) \mathcal{E}^{(\theta',g)}(1-z) \,.
\end{aligned}
\end{equation}
We conclude this appendix by noticing that the procedure to obtain equation \eqref{eq:almostourconn} has some ambiguities. In particular, the continuation of the $\gamma$ contour to obtain equation \eqref{eq:newcont} is not unique, and different choices give different answer. Moreover, since all the crossing constraints are only sensitive to the modulus squared of the conformal blocks, there are phase ambiguities. In \cite{Bonelli:2022ten} the phase ambiguities were fixed by matching the connection coefficients in some regions where the ODE is approximated by some hypergeometrics. Here we fix the various ambiguities by comparing with the WKB answer in the appropriate regime, as discussed in the main text.

\section{The extremal BTZ correlator}\label{app:BTZ}
We now compute the thermal two point function in the extremal BTZ background. Note that in presence of a horizon the prescription to compute boundary correlator is quite different. We refer to \cite{Son:2002sd} for an exhaustive discussion.

The wave equation in the metric \eqref{eq:standardbtz} (we set $a_0 = \tilde{n} = 1$ without loss of generality) reduces to
\begin{equation}
    \partial_z^2 \psi(z) + \left(-\frac{\Delta \left(\Delta-2\right)}{4 z^2} + i \frac{\ell+\omega}{4 z} - \frac{1}{4} \right) \psi(z) = 0 \,, 
\label{eq:radwave}
\end{equation}
with $z=i\frac{\ell-\omega}{r^2}$ and for a perturbation of the form
\begin{equation}
\phi\left(r, \tau, \sigma\right) = \sum_{\ell} \int d \omega \, e^{i \tau \omega + i \sigma \ell} \psi(r) \,.
\label{eq:btzwave}
\end{equation}
Equation \eqref{eq:btzwave} has a regular singularity at $z=0$ and an irregular singularity at $z = \infty$. It is the confluent hypergeometric equation. Close to $z \sim \infty$ the two linear independent solutions are given by
\begin{equation}
W_{\frac{ip}{2}, \frac{\Delta-1}{2}} \left(z\right) \,, \quad W_{-\frac{ip}{2}, \frac{\Delta-1}{2}} \left(e^{-i \pi}z\right) \,.
\end{equation}
where $p, w$ are the lightcone momenta introduced in \eqref{eq:lightconem}. $W_{\mu, \kappa}(z)$ is the $W$ Whittaker function. Near $z \sim \infty$
\begin{equation}
W_{\mu, \kappa}(z) \simeq e^{-\frac{z}{2}} z^{\mu}\left(1+\mathcal{O}\left(z^{-1}\right)\right) \,.
\end{equation}
In order to compute the retarded Green's function in momentum space we impose the boundary condition that the wave is purely ingoing into the horizon, that is
\begin{equation}
    \psi(z) = W_{-\frac{ip}{2}, \frac{\Delta-1}{2}} \left(-z\right) \,.
\label{eq:ingoingbtz}
\end{equation}
The solution of the connection problem for the Whittaker functions is well known. Expanding $\psi(z)$ close to the AdS boundary ($z\to  0$) gives 
\begin{equation}
    \psi(z) = \frac{\Gamma\left(\Delta-1\right)}{\Gamma\left(\frac{\Delta + i p}{2}\right)} M_{\frac{ip}{2},\frac{-\Delta+1}{2}}(z) + \frac{\Gamma\left(1-\Delta\right)}{\Gamma\left(1+\frac{i p - \Delta}{2}\right)} M_{\frac{ip}{2},\frac{\Delta-1}{2}}(z)
\label{eq:psiz}
\end{equation}
where
\begin{equation}
M_{\mu, \kappa} (z) = e^{-\frac{1}{2}z} z^{\frac{1}{2}+\kappa} \sum_{n\ge0} \frac{\left(\frac{1}{2}+\kappa-\mu\right)_n}{\left(1+2\kappa\right)_n n!} z^n \,,
\end{equation}
where $(a)_n$ is the rising Pochhammer symbol. The retarded Green's function is given to be the ratio of the coefficient of $r^{-\Delta}$ and of $r^{\Delta-2}$, that is
\begin{equation}
\begin{aligned}
 G_R &= \left(i (\ell-\omega)\right)^{\Delta - 1} \frac{\Gamma\left(1-\Delta\right) \Gamma\left(\frac{1}{2} \left(\Delta+i \frac{\ell+\omega}{2}\right)\right)}{\Gamma\left(\Delta-1\right) \Gamma\left(1-\frac{1}{2} \left(\Delta-i \frac{\ell+\omega}{2}\right)\right)} = \\ &= \left(2i w\right)^{\Delta - 1} \frac{\Gamma\left(1-\Delta\right) \Gamma\left(\frac{1}{2} \left(\Delta+i p\right)\right)}{\Gamma\left(\Delta-1\right) \Gamma\left(1-\frac{1}{2} \left(\Delta-i p\right)\right)} \,.
\label{eq:RBTZmomentum}
\end{aligned}
\end{equation}
This expression has poles at
\begin{equation}
\omega_n = -\ell + 2i (2n+\Delta) \,.
\end{equation}
Note that the poles are all in the upper half plane, as a consequence of the convention for the Fourier transform given in \eqref{eq:btzwave}. Feynman and Wightman correlators will respectively be given by
\begin{equation}
\begin{aligned}
G_W &= \text{sign} \,  \omega \, \text{Im} G_R \,, \\
G_F &= \text{Re} G_R + i \, \text{sign} \, \omega \,  \text{Im} G_R \,.
\end{aligned}
\end{equation}
For concreteness,
\begin{equation}
\text{Im} G_R(w,p) = \frac{1}{2i}\frac{\Gamma\left(1-\Delta\right)}{\Gamma\left(\Delta-1\right)} \left(\left(2i w\right)^{\Delta - 1}\frac{\Gamma\left(\frac{\Delta-ip}{2}\right)}{\Gamma\left(1-\frac{\Delta+ip}{2}\right)} - \left(-2i w\right)^{\Delta - 1}\frac{\Gamma\left(\frac{\Delta+ip}{2}\right)}{\Gamma\left(1-\frac{\Delta-ip}{2}\right)}\right) \,.
\label{eq:imgrbtz}
\end{equation}

It is interesting to compare the BTZ correlator with the small $\eta$ limit of the correlator computed in the pure heavy state \eqref{defOH}. As outlined in the main text (see equation \eqref{eq:Gimw}), the correlator in the $\tilde{n}=1$  background reduces to \eqref{eq:RBTZmomentum} when $w$ gets a negative imaginary part and $\eta\to 0$. The idea is that when $w$ is in the lower half plane, imposing regularity at $\rho=0$ is approximatively the same as selecting the purely ingoing wave in the $AdS_2$ throat region,  which for $\eta=0$ becomes the near horizon region of an extremal BTZ black hole. The relevant regime for this discussions is $\rho\ll\eta^{-1}$ with $\eta \ll 1$, since it captures both the $\rho \sim 0$ region and the $1\ll\rho\ll\eta^{-1}$ $AdS_2$ throat region. It is convenient to rewrite the wave equation \eqref{eq:schrodform} with potential \eqref{eq:RCHE} in the coordinate $x=\eta^{-2} z$. When $\rho \sim 0$, $x \sim 0$, and when $\rho$ is in the $AdS_2$ region $x \sim \infty$. The small $\eta$ potential reads
\begin{equation}
V(x) \simeq \frac{w^2}{x \eta^2} + \frac{1-(p+w)^2-4pwx-4w^2 x}{4x^2} \,.
\end{equation}
The corresponding wave equation is solved in terms of Bessel functions. Imposing regularity at $x=0$ we find
\begin{equation}
\psi(x) \simeq \left(\frac{iw}{\eta^2}\right)^{-p-w} \sqrt{\eta^2 x} I_{p+w}\left(\frac{2iw \sqrt{x}}{\eta}\right) \,,
\end{equation}
where $I_\nu(z)$ is a modified Bessel function of the first kind. For $1\ll\rho\ll\eta^{-1}$ the regular solution looks like
\begin{equation}
\psi(x) \simeq \frac{\Gamma\left(1+p+w\right)\eta^{\frac{3}{2}} (-x)^{\frac{1}{4}}}{2\sqrt{\pi w}} \left( \frac{iw}{\eta^2} \right)^{-p-w} \left( -i e^{i\frac{2w\sqrt{x}}{\eta}} + e^{i\pi(p+w)} e^{-i\frac{2w\sqrt{x}}{\eta}} \right) \,.
\end{equation}
In this regime $x \simeq \eta^{-2}-(\eta \rho)^{-2}$, and in terms of $\rho$ we have, ignoring overall factors,
\begin{equation}
\begin{aligned}
&\psi(x) \sim c_1 \, e^{-i \frac{w}{\eta^2\rho^2}} + c_2 \, e^{i \frac{w}{\eta^2\rho^2}} \,, \\
&c_1 = -i \, e^{\frac{2iw}{\eta^2}} \quad , \quad c_2 = e^{i\pi (p+w)} e^{-\frac{2iw}{\eta^2}} \,.
\end{aligned}
\end{equation}
When $w = w_R-i w_I$ with $w_I>0$, $c_2$ is exponentially suppressed with respect to $c_1$,  and accordingly
\begin{equation}
\psi(x) \sim c_1 \, e^{-i \frac{w}{\eta^2\rho^2}} + \mathcal{O}\left(e^{-\frac{2w_I}{\eta^2}}\right) \,.
\end{equation} 
Comparing with \eqref{eq:ingoingbtz} we see that this corresponds to a purely ingoing wave in the $AdS_2$ throat.

\section{WKB analysis}
\label{sec:wkb}

In the semiclassical limit of large frequency, $\omega$, the wave equation, and hence the propagator $G(\omega,\ell)$,  can be analytically evaluated using the WKB method, an approach that has already been applied to this problem in \cite{Bena:2019azk}. Here we reconsider this approach mainly as a check of our exact formula \eqref{eq:largeLG} for the $\tilde n=1$ propagator in the small $\eta$ limit. In particular we focus on the prediction for the real poles, which we have derived from \eqref{eq:largeLG} in \eqref{eq:realpoleslargeL}. The WKB computation for general values of $\omega$ and $\ell$ is quite cumbersome and, hence, we restrict to two particular cases, $\ell=0$ and $p=0$, that enjoy significant simplifications and still allow us to test many details of the general formula \eqref{eq:largeLG}.

\subsection{$\ell=0$}
When $\ell=0$ the wave equation simplifies to 
\begin{equation}
\begin{aligned}
&\left(\partial_z^2 + k^2_\omega (z)\right) u(z) = 0 \,, \\
&k^2_\omega(z) = - \omega^2 \frac{1-\eta^2}{4 \eta^4} \frac{z-z_t}{z(1-z)}+\frac{1}{4 z^2} - \frac{\Delta(\Delta-2)}{4z(1-z)^2} \quad \mathrm{with}\quad z_t = \frac{1}{1-\eta^2}\,.
\end{aligned}
\end{equation}
The potential is controlled by the same parameter $L^2 =- \omega^2 \frac{1-\eta^2}{\eta^4}$ which plays an important role for the exact propagator $G(\omega,0)$ derived in Section~\ref{sec:ntilde1exact}. In the limit of large $L^2$ -- which can be attained by taking large $\omega$ and/or small $\eta$ -- and finite $\Delta$ the potential can be further simplified to
\begin{equation}
  k^2_\omega(z) \approx - \omega^2 \frac{1-\eta^2}{4 \eta^4} \frac{z-z_t}{z(1-z)}  \,,
  \label{eq:wkb}
\end{equation}
and the wave equation is well approximated by a WKB expansion. Note however that this approximation fails close to $z=0$ and $z=1$ where the quadratically divergent terms, $\sim z^{-2}$ or $\sim (1-z)^{-2}$, of the potential dominate. We will come back to this point in due course.

The only turning point of the wave equation, where $k_\omega$ vanishes, is 
$z_t = \frac{1}{1-\eta^2}$, and it always lies beyond the physical region  
$z \in [0,1]$  for the allowed values of $\eta \in (0,1)$. Thus, in the whole physical region the WKB wave function reads
\begin{equation}\label{eq:uwkb}
u(z) = \frac{1}{\sqrt{k_\omega}} \left(A e^{\mathcal{Z}(0,z)} + B e^{-\mathcal{Z}(0,z)}\right) =  \frac{1}{\sqrt{k_\omega}} \left(A' e^{\mathcal{Z}(1,z)} + B' e^{-\mathcal{Z}(1,z)}\right) \,,
\end{equation}
where we have defined
\begin{equation}
    \mathcal{Z}(z_1,z_2) = i \int_{z_1}^{z_2}\!dz\, k_\omega (z)=i |\omega|\frac{\sqrt{1-\eta^2}}{2\eta^2}\int_{z_1}^{z_2}\!dz\,\sqrt{\frac{z_t-z}{z(1-z)}}\,,
\end{equation}
and
\begin{equation}
A' = A \, e^{\mathcal{Z}(0,1)} \,, \quad B' = B \, e^{-\mathcal{Z}(0,1)} \,.
\end{equation}
For definiteness we will first consider the case $\omega>0$. Close to $z=0$ the above wave function is a linear combination of the two oscillating terms $e^{\pm \mathcal{Z}(0,z)} \simeq e^{\pm i \frac{\omega \sqrt{z}}{\eta^2}}$ and it is not obvious to pick the linear combination corresponding to the regular solution. This is a consequence of the aforementioned failure of the WKB approximation near $z\approx 0$. To correctly describe the wave function near the origin and at the same time to have a non-trivial overlap with the WKB region, one should include in the potential both the quadratically and the linearly divergent terms, $\sim z^{-2}$ and $z^{-1}$; hence one should approximate 
\begin{equation}
k^2_\omega(z) \approx \left( \frac{\omega^2}{4\eta^2 z} +\frac{1}{4z^2} - \frac{\Delta(\Delta-2)}{4z} \right) \,.
\end{equation}
The regular solution of the wave equation with this potential is 
\begin{equation}
u_{reg} (z) = 2 i \sqrt{\pi z} I_0 \left(i \sqrt{\frac{\omega^2-\Delta(\Delta-2)\eta^4}{\eta^4} z}\right) \,.
\end{equation}
To match with the WKB wave function we expand the previous expression for large $\omega$ and $z$ not too close to $0$, such that $\omega\sqrt{z}\gg 1$: 
\begin{equation}\label{eq:ureg}
u_{reg} (z)\approx 2 i \sqrt{\frac{2 \eta^2 \sqrt{z}}{\omega}} \sin \left(\frac{\omega \sqrt{z}}{\eta^2} + \frac{\pi}{4}\right) \,.
\end{equation}
Comparing \eqref{eq:ureg} and \eqref{eq:uwkb} fixes
\begin{equation}
A = e^{i \frac{\pi}{4}} \,, \quad B = - e^{-i \frac{\pi}{4}} \,.
\end{equation}
An analogous matching procedure has to be performed in the asymptotic region, $z\approx 1$, to identify the normalizable and non-normalizable solutions, since in \eqref{eq:wkb} we are also neglecting a term that diverges quadratically as $z \to 1$. Proceeding as before we find that, in the region that overlaps with the WKB wave function, the normalizable and non normalizable solutions are
\begin{equation}
\begin{aligned}
&u_{norm} (z) = c \, 2 i \sqrt{\frac{2 \eta \sqrt{1-z}}{\omega}} \sin \left(- \frac{\omega \sqrt{1-z}}{\eta} + \frac{\pi}{4} (2\Delta+1)\right) \,, \\
&u_{n-norm} (z) = \tilde{c} \, 2 \sqrt{\frac{2 \eta \sqrt{1-z}}{\omega} }\cos \left( \frac{\omega \sqrt{1-z}}{\eta} + \frac{\pi}{4}(2\Delta+1) \right) \,.
\end{aligned}
\end{equation}
We choose the normalizations 
\begin{equation}
\begin{aligned}
&c = - i 2^{\Delta-2} (\pi)^{-\frac{1}{2}} \left(\omega/\eta\right)^{1-\Delta} \Gamma\left(\Delta\right) \,, \\
&\tilde{c} = - 2^{-\Delta} (\pi)^{-\frac{1}{2}} \left(\omega/\eta\right)^{\Delta-1}\Gamma\left(2-\Delta\right) \,, \\
\end{aligned}
\end{equation}
so that at leading order close to $z\approx 1$ the two solutions go as $(1-z)^{\frac{1}{2}\pm a_1}$. Finally, we can rewrite the semiclassical wave function close to $z \approx 1$ in terms of $u_{norm}$ and $ u_{n-norm}$:
\begin{equation}
\begin{aligned}
u(z) =  - \frac{2}{\sin \pi \Delta} \bigg[ & \left(A' e^{i \frac{\pi}{4}(1+2\Delta)} - B' e^{- i \frac{\pi}{4}(1+2\Delta)} \right) \frac{u_{norm}(z)}{c} + \\ &+ \left(A' e^{i \frac{\pi}{4}(1+2\Delta)} + B' e^{i \frac{\pi}{4}(1+2\Delta)} \right) \frac{u_{n-norm}(z)}{\tilde{c}} \bigg] .
\label{eq:wkbsolnApBp}
\end{aligned}
\end{equation}
Following the usual prescription, the correlator reads
\begin{equation}
G(\omega, 0) = \frac{\tilde{c}}{c} \frac{A' e^{i \frac{\pi}{4}(2\Delta+1)}-B' e^{-i \frac{\pi}{4}(2\Delta+1)}}{A' e^{-i\frac{\pi}{4}(2\Delta+1)}+B' e^{i \frac{\pi}{4}(2\Delta+1)}} = - e^{i \pi \Delta} \left(\frac{4 \omega^2}{\eta^2}\right)^{\Delta-1} \frac{1+ie^{-i \pi\Delta} B'/A' }{1+ i e^{i \pi \Delta} B'/A'}
\label{eq:corrABAB}
\end{equation}
Substituting the values of $A', B'$ found before we have
\begin{equation}
G(\omega, 0) = -e^{-i\pi\Delta} \left(\frac{\omega^2}{4\eta^2}\right)^{\Delta-1} \frac{\Gamma\left(2-\Delta\right)}{\Gamma\left(\Delta\right)} \frac{1-e^{2\mathcal{Z}(0,1)+i\pi\Delta}}{1-e^{2\mathcal{Z}(0,1)-i\pi\Delta}}  \quad \text{for} \quad \omega > 0 \,.
\label{eq:Gwkbposom}
\end{equation}
Note that this expression is real for real $\omega$ as it should. We stress that this  is only valid for $\omega>0$. However, since the $\ell = 0$ wave equation is invariant under $\omega \to - \omega$ we can readily obtain the negative $\omega$ result by just reflecting the frequency in \eqref{eq:Gwkbposom}:
\begin{equation}
G(\omega, 0) = -e^{i\pi\Delta} \left(\frac{\omega^2}{4\eta^2}\right)^{\Delta-1} \frac{\Gamma\left(2-\Delta\right)}{\Gamma\left(\Delta\right)} \frac{1-e^{2\mathcal{Z}(0,1)-i\pi\Delta}}{1-e^{2\mathcal{Z}(0,1)+i\pi\Delta}} \,, \quad \text{if} \quad \omega < 0 \,.
\label{eq:Gwkbnegom}
\end{equation}
Equation \eqref{eq:Gwkbposom} has poles when
\begin{equation}
2 \mathcal{Z}(0,1) = i \pi \Delta + 2 \pi i n \,.
\end{equation}
Since
\begin{equation}
\mathcal{Z}(0,1) = \frac{i \omega}{\eta^2} E(1-\eta^2)
\end{equation}
where $E$ is the complete elliptic integral of the second kind, we find
\begin{equation}
\omega_n = \frac{\pi \eta^2}{2 E(1-\eta^2)} (2n+\Delta) = 
\begin{cases}
(2n+\Delta)\left(1-\frac{3}{4}(1-\eta^2) + \mathcal{O}\left((1-\eta^2)^4\right) \right) \,, \quad \eta^2 \approx 1 \,, \\
\frac{\pi \eta^2}{2}(2n+\Delta)\left(1+\frac{1}{4}\left(1+\log \frac{\eta^2}{16}\right)\eta^2 + \dots \right) \,, \quad \eta^2 \approx 0 \,.
\end{cases}
\label{eq:polesWKBell0}
\end{equation}
The $\eta^2 \approx 1$ result is in agreement with the large $n$ limit of what we found in \eqref{eq:realpolesh1}. When $\eta^2 \to 0$ the spacing between the poles goes to zero as $\eta^2$ as for the $\tilde{n} =0$ case. Moreover the position of the poles matches with the result \eqref{eq:realpoleslargeL}, derived from the exact propagator \eqref{eq:largeLG} in the large $L$ regime. 

\subsection{$p=0$}
One can perform an analogous WKB computation with $p=0$. For simplicity, and also because it will serve us as a further check of Eq. \eqref{eq:realpoleslargeL}, we will restrict to the derivation of the poles of the propagator, which correspond to normalizable wave functions. The potential for $p=0$ is
\begin{equation}
 k^2_w(z)=w^2 \frac{1-\eta^2}{\eta^4} \frac{z-z_t}{z^2} + \frac{1}{4z^2} -\frac{\Delta(\Delta-2)}{4z(1-z)^2} \quad\mathrm{with}\quad z_t=\frac{\eta^4}{4(1-\eta^2)}\,.   
\end{equation}
The potential is again controlled by $L^2=-4 w^2 \frac{1-\eta^2}{\eta^4}$, and if we take $L$ large and $\Delta$ finite we can approximate
\begin{equation}
k^2_w(z)\approx w^2 \frac{1-\eta^2}{\eta^4} \frac{z-z_t}{z^2}
\end{equation}
everywhere (including $z\approx 0$) but the asymptotic region close to $z=1$, which will be dealt with separately. We focus on the $\eta\to 0$, in which the turning point $z_t$ lies within the physical range $z\in [0,1]$. Then the WKB wave function has to be defined in two patches, $(0,z_t)$ and $(z_t,1)$, where it is given, respectively, by
\begin{equation}
u_<(z)= \frac{1}{\sqrt{k_w(z)}} \left[ A e^{-\mathcal{Z}(z_t,z)} + B e^{\mathcal{Z}(z_t,z)} \right]\quad , \quad z\in (0,z_t)\,,
\end{equation}
\begin{equation}
u_>(z)= \frac{1}{\sqrt{k_w(z)}} \left[ C e^{\mathcal{Z}(z_t,z)} + D e^{-\mathcal{Z}(z_t,z)} \right]\quad ,\quad z\in (z_t,1)\,,
\end{equation}
with
\begin{equation}
 \mathcal{Z}(z_t,z) = i \int_{z_t}^z \!dx\,k_w(x)\approx |w| \frac{i\sqrt{1-\eta^2}}{\eta^2} \int_{z_t}^z\!dx\,\frac{\sqrt{x-z_t}}{x} \,.
\end{equation}
We consider $w>0$ in the following. Then $\mathcal{Z}(z_t,z)\to +\infty$ in the limit $z\to $ and the regular solution is the one with $B=0$. The standard crossing formulas give
\begin{equation}
    C = A \,e^{-i \frac{\pi}{4}}\quad ,\quad  D = A \,e^{i \frac{\pi}{4}}\,.
\end{equation}
The semiclassical wave function $u_>$ is then a linear combination of oscillating terms, which has to be matched with the normalizable solution at $z\approx 1$, where the WKB approximation fails. In the region close to $z\approx 1$ one can approximate
\begin{equation}
   k^2_w(z)\approx -\frac{\Delta(\Delta-2)}{4(1-z)^2}-\frac{\Delta(\Delta-2)-1}{4(1-z)}-\frac{1}{4}\left[\Delta(\Delta-2)-2 +w^2 \left(1-4\frac{1-\eta^2}{\eta^4}\right)\right]\,,
\end{equation}
and then the wave equation is solved by Whitaker functions. One can show that the normalizable solution expanded for large $w(1-z)$ matches the WKB wave function $u_>$ if
\begin{equation}
2 \mathcal{Z}(z_t,1) = i\,\pi\,\left(\frac{(\Delta+1)}{2} + 2 n \right)\quad , \quad n\in \mathbb{Z}\,.
\end{equation}
This is the condition that determines the poles of the propagator. Evaluating $\mathcal{Z}(z_t,1)$ for $\eta\to 0$ leads to the equation
  \begin{equation}
\frac{4 w}{\eta^2} -(2+\pi) w +O(\eta^2) = \pi\,\left(2n+\frac{(\Delta+1)}{2}\right)\quad \mathrm{for}\quad  w>0 \,,
\end{equation}  
and thus the positive poles at $p=0$ are
\begin{equation}
w_n=\frac{\pi\eta^2}{2}
\left(n+\frac{(\Delta+1)}{4}\right)
\left[1+\frac{\eta^2}{4}(2+\pi)+O(\eta^4)\right]\,,\quad n>0\,.
\label{eq:polesWKBp0}
\end{equation}
We again find agreement with \eqref{eq:realpoleslargeL}.



\providecommand{\href}[2]{#2}\begingroup\raggedright\endgroup

\end{document}